\documentclass[11pt,a4paper]{article}
\usepackage{jheppub}
\usepackage{graphicx}
\usepackage{cancel}
\usepackage{caption}
\usepackage{placeins}
\usepackage{bm}
\usepackage{hyperref}
\usepackage{amsmath}
\usepackage{amssymb}
\usepackage{listings}
\usepackage{axodraw4j}
\usepackage[toc, page]{appendix}
\usepackage{color}
\newcommand{\order}[1]{{\cal O}(#1)}

\newcommand{\amc}{{\sc MadGraph5\_aMC@NLO}}

\newcommand{\msbar}{\overline{\mbox{\small MS}}}

\title{Anatomy of double heavy-quark initiated processes}
\author[a]{Matthew Lim,} 
\author[b]{Fabio Maltoni,}
\author[c]{Giovanni Ridolfi,}
\author[a]{Maria Ubiali}
\affiliation[a]{Cavendish Laboratory, J.J. Thomson Avenue, Cambridge, UK}
\affiliation[b]{Centre for Cosmology, Particle Physics and Phenomenology CP3,
Universit\'{e} Catholique de Louvain, Chemin du Cyclotron,
1348 Louvain--la--Neuve, Belgium}
\affiliation[c]{Dipartimento di Fisica, Universit\`a di Genova 
\& INFN, Sezione di Genova \\
Via Dodecaneso 33, 16146 Genova, Italy}
\emailAdd{malim@hep.phy.cam.ac.uk}
\emailAdd{fabio.maltoni@uclouvain.be}
\emailAdd{giovanni.ridolfi@ge.infn.it}
\emailAdd{ubiali@hep.phy.cam.ac.uk}

\abstract{A number of phenomenologically relevant processes at hadron
  colliders, such as Higgs and $Z$ boson production in association
  with $b$ quarks, can be conveniently described as scattering of
  heavy quarks in the initial state.  We present a detailed analysis
  of this class of processes, identifying the form of the leading
  initial-state collinear logarithms that allow the relation of
  calculations performed in different flavour schemes in a simple and
  reliable way.  This procedure makes it possible to assess the size
  of the logarithmically enhanced terms and the effects of their
  resummation via heavy-quark parton distribution functions. As an
  application, we compare the production of (SM-like and heavy) scalar
  and vector bosons in association with $b$ quarks at the LHC in the
  four- and five-flavour schemes as well as the production of a heavy $Z'$
  in association with top quarks at a future 100 TeV hadron collider
  in the five- and six-flavour schemes.  We find that, in agreement with a
  previous analysis of single heavy-quark initiated processes, the
  size of the initial-state logarithms is mitigated by a kinematical
  suppression. The most important effects of the resummation are a
  shift of the central predictions typically of about 20\% at a
  justified value of the scale of each considered process and a
  significant reduction of scale variation uncertainties.  }

\arxivnumber{1604.xxxxx}
\subheader{\small CAVENDISH-HEP-16/07, CP3-16-24 }
\begin{document}

\maketitle
\flushbottom

\section{Introduction}

With the imminent restart of data-taking at LHC Run II the need for
accurate theoretical predictions for energetic final states, typically
involving the production of heaviest particles of the Standard Model
(SM), becomes more and more pressing. The study of associated
production of (possibly new) vector or scalar bosons in association
with heavy quarks, such as top and bottom quarks, are among the
highest priorities of the new run.  In particular, $b$ quarks play an
important role in the quest for new physics as well as for precise SM
measurements from both an experimental and a theoretical
perspective. Firstly, they provide a very clean signature as they may
easily be identified in a detector due to the displacement of vertices
with respect to the collision point, a consequence of the $b$-quark
long lifetime. Secondly, the relative strength of the Higgs Yukawa
coupling (or possibly of new scalar states) to the heavy quarks is
important in determining the phenomenology, both in production as well
as in decay. In particular, production associated with $b$ quarks
could provide the leading mode for Higgs bosons with enhanced Yukawa
couplings in many scenarios beyond the Standard Model.

At hadron colliders, any process that features heavy quarks can be
described according to two different and complementary approaches.  In
the {\it massive} or four-flavor (4F) scheme (in the case of $b$
quarks), the heavy quark is produced in the hard scattering and arises
as a massive particle in the final state.  The dependence on the heavy
quark mass $m_b$ is retained in the matrix element and explicit
logarithms of $Q/m_b$, $Q$ being some hard scale of the process,
appear at each order in perturbation theory as a result of collinearly
enhanced (yet finite) splittings $q \to q g$ or of a gluon into heavy
quark pairs, $g \to q \bar q$.  On the other hand, in the {\it
  massless} or five-flavour (5F) scheme (in the case of $b$ quarks),
$Q\gg m_b$ is assumed and the heavy quark is treated on the same
footing as the light quarks: it contributes to the proton wave
function and enters the running of the strong coupling constant
$\alpha_s$. In this scheme the heavy quark mass is neglected in the
matrix element and the collinear logarithms that may spoil the
convergence of the perturbative expansion of the 4F scheme cross
section are resummed to all orders in the evolution of the heavy quark
parton density.

In a previous work~\cite{Maltoni:2012pa}, we examined processes
involving a single $b$ quark in both lepton-hadron and hadron-hadron
collisions. It was found that, at the LHC, unless a very heavy
particle is produced in the final state, the effects of initial-state
collinear logarithms are always modest and such logarithms do not
spoil the convergence of perturbation theory in 4F scheme
calculations. This behaviour was explained by two main reasons, one of
dynamical and the other of kinematical nature.  The first is that the
effects of the resummation of the initial-state collinear logarithms
is relevant mainly at large Bjorken-$x$ and in general keeping only
the explicit logs appearing at NLO is a very good approximation. The
second reason is that the na{\"i}ve scale $Q$ that appears in the
collinear logarithms turns out to be suppressed by universal phase
space factors that, at hadron colliders, reduce the size of the
logarithms for processes taking place.  As a result, a consistent and
quantitative analysis of many processes involving one $b$ quark in the
initial state was performed and a substantial agreement between total
cross sections obtained at NLO (and beyond) in the two schemes found
within the expected uncertainties.

In this work we focus on processes that can be described by two $b$
quarks in the initial state, such as $pp\to H b\bar{b}$ or $pp\to Z
b\bar{b}$. As already sketched in ~\cite{Maltoni:2012pa}, the same
arguments used for single heavy-quark initiated processes can be used
to analyse the double heavy-quark case. One may na{\"i}vely expect
that the resummation effects for processes with two $b$ quarks in the
initial state can be simply obtained by ``squaring'', in some sense,
those of processes with only one $b$ quark. There are, however, a
number of features that are particular to the double heavy-quark
processes and call for a dedicated work.  One is that the lowest order
contribution in the 4F scheme appears for the first time among the
NNLO real corrections to the leading order 5F scheme
calculation. Furthermore, due to the simplicity of the 5F description
(i.e. Born amplitudes are $2 \to 1$ processes), results in the 5F scheme are
now available at NNLO, while, thanks to the progress in the automation
of NLO computations, 4F scheme results have become easily accessible for a
wide range of final states.  In fact, it is easy to understand that a
meaningful comparison between the two schemes for double heavy-quark
initiated processes starts to be accurate if results are taken at NNLO
for the 5F and at NLO for the 4F case.

Both $pp\to H b\bar{b}$ or $pp\to Z b\bar{b}$ have been considered in
previous works.  For the LHC, it was demonstrated that consistent
results for both the total cross section and differential
distributions for bottom-fusion initiated Higgs production can be
obtained in both
schemes~\cite{Dawson:2003kb,Maltoni:2003pn,Campbell:2004pu,Harlander:2011aa,Wiesemann:2014ioa}.
Analogous studies were performed for bottom-fusion initiated $Z$
production~\cite{Campbell:2000bg,Maltoni:2003pn,FebresCordero:2008ci,Cordero:2009kv,Frederix:2011qg}.
All these studies suggested that the appropriate factorisation and
renormalisation scales associated to these processes are to be chosen
smaller than the mass of the final state heavy particles. In
particular, scales of about $M_{H,Z}/4$ have been proposed in order to
stabilise the perturbative series and make the four- and five-flavor
predictions closer other.  $(M_H+2m_b)/4$ is the scale adopted by the
LHC Higgs Cross Section Working Group (HXSWG) to match the NLO 4F and
NNLO 5F scheme predictions in case of bottom-fusion initiated Higgs
production via the Santander interpolation~\cite{Harlander:2011aa} and
via the use of consistently matched
calculations~\cite{Bonvini:2015pxa,Forte:2015hba,Bonvini:2016fgf}.

While previous studies support a posteriori the evidence that smaller
scales make the four- and five-flavor pictures more consistent, no complete
analysis of the relation of the two schemes in the case of double
heavy-quark initiated processes has been provided. In particular, no
analytic study of the collinear enhancement of the cross section and
the kinematics of this class of processes has been performed.

In this work, we fill this gap by extending our previous work to
double heavy-quark production. We first present an analytic comparison
of the two schemes that allow us to unveil a clear relation between
them, establish the form of the logarithmic enhancements and determine
their size. We then compare the predictions for LHC phenomenology in a
number of relevant cases focusing on LHC Run II.  Furthermore, we
expand our investigation to high energy processes involving top quarks
at future colliders.  At centre-of-mass energies of order $100$~TeV, a
new territory far beyond the reach of the LHC would be explored. At
such an energy, much heavier particles could be produced at colliders and
top-quark PDFs may become of relevance in processes involving top
quarks in the initial state.

The structure of the work is as follows. In Sect.~\ref{sec:analytic}
we examine the kinematics of 2 to 3 body scattering and calculate the
phase space factor for the particular case of $b$-initiated Higgs
production---we thus derive the logarithmic contributions to the cross
section which arise in a 4F scheme. We then proceed to generate
kinematic distributions for the processes and use these to analyse the
4F and 5F scheme results. We conclude the section by suggesting a
factorisation scale at which results from either process may be
meaningfully compared. In Sect.~\ref{sec:numeric} we compare the
results on total cross sections obtained in both schemes for a number
of phenomenologically relevant processes at the LHC and future
colliders.  Finally, our conclusions are presented in
Sect.~\ref{sec:conclusions}.

\section{Different heavy quark schemes: analytical comparison}
\label{sec:analytic}
We start by considering Higgs boson production via $b\bar b$ fusion
in the 4F scheme. The relevant partonic subprocess is
\begin{equation}
g(p_1)+g(p_2)\to b(k_1)+H(k)+\bar{b}(k_2),
\label{ggbbH}
\end{equation}
where the $b$ quarks in the final state are treated as massive
objects.  Since the $b$-quark mass $m_b$ is much smaller than the
Higgs boson mass $M_H$, we expect the cross section for the process
(\ref{ggbbH}) to be dominated by the configurations in which the two
final-state $b$ quarks are emitted collinearly with the incident
gluons. Indeed the quark-antiquark channel ($q\bar{q}\to b\bar{b}H$)
that also contributes to the leading-order cross section in the 4F
scheme is very much suppressed with respect to the gluon-gluon one. In
order to estimate the importance of large transverse momentum $b$
quarks in the gg channel, as compared to the dominant collinear
configurations, we will perform an approximate calculation of the
cross section for the process (\ref{ggbbH}) limiting ourselves to the
dominant terms as $m_b\to 0$. The result will then be compared to the
full leading-order 4F scheme calculation.  We present here the final
result; the details of the calculation can be found in
Appendix~\ref{appxs}.

The differential partonic cross section can be expressed as a function of
five independent invariants, which we choose to be
\begin{equation}
\hat s=(p_1+p_2)^2;\quad
t_1 = (p_1-k_1)^2;\quad
t_2 = (p_2-k_2)^2;\quad
s_1 = (k_1+k)^2 ;\quad
s_2 = (k_2+k)^2 .
\end{equation}
Collinear singularities appear, for $m_b^2=0$, either when
\begin{equation}
t_1\to 0;\qquad t_2\to 0,
\label{collt}
\end{equation}
or when
\begin{equation}
u_1\to 0;\qquad u_2\to 0,
\label{collu}
\end{equation}
where
\begin{equation}
u_1=(p_1-k_2)^2;\qquad u_2=(p_2-k_1)^2.
\end{equation}
The configuration in Eq.~(\ref{collt}) is achieved for 
\begin{equation}
k_1=(1-z_1)p_1;\qquad k_2=(1-z_2)p_2;\qquad 0\leq z_i\leq 1
\end{equation}
while the one in Eq.~(\ref{collu}) corresponds to
\begin{equation}
k_1=(1-z_1)p_2;\qquad k_2=(1-z_2)p_1.
\end{equation}
In both cases we find
\begin{equation}
\hat s=\frac{M_H^2}{z_1z_2};\qquad 
s_1=\frac{M_H^2}{z_1};\qquad s_2=\frac{M_H^2}{z_2}.
\end{equation}
An explicit calculation yields
\begin{equation}
\hat\sigma^{\rm 4F,coll}(\hat\tau)
=\hat\tau\frac{\alpha_s^2}{4\pi^2}
\frac{G_F \pi}{3\sqrt{2}}\frac{m_b^2}{M_H^2}
2\int_0^1 dz_1\int_0^1 dz_2\,
P_{qg}(z_1)P_{qg}(z_2)
L(z_1,\hat\tau) L(z_2,\hat\tau) 
\delta\left(z_1z_2-\hat\tau\right),
\label{parton4F1}
\end{equation}
where
\begin{equation}
\hat\tau=\frac{M_H^2}{\hat s},
\end{equation}
$P_{qg}(z)$ is the leading-order quark-gluon
Altarelli-Parisi splitting function
\begin{equation} 
P_{qg}(z) = \frac{1}{2} [z^2+(1-z)^2],
\label{Pqg}
\end{equation} 
and
\begin{equation}
L(z,\hat\tau)=\log\left[\frac{M_H^2}{m_b^2}\frac{(1-z)^2}{\hat\tau}\right].
\label{logs}
\end{equation}
The suffix ``coll'' reminds us that we are neglecting less singular
contributions as $m_b\to 0$, i.e.\ either terms with only one
collinear emission, which diverge as $\log m_b^2$, or terms which are
regular as $m_b\to 0$.

We now observe that the leading-order partonic cross section for the process
\begin{equation}
b(q_1)+\bar{b}(q_2)\to H(k),
\end{equation}
relevant for calculations in the 5F scheme, is given by~\cite{stirling}
\begin{equation}
\hat\sigma^{\rm 5F}(\hat\tau) =\frac{G_F\pi}{3\sqrt{2}}\frac{m_b^2}{M_H^2}
\delta(1-\hat\tau),
\label{parton5F}
\end{equation} 
with
\begin{equation}
\hat s=(q_1+q_2)^2.
\end{equation}
Hence, the 4F scheme cross section in the collinear limit, Eq.~(\ref{parton4F1}),
can be rewritten as
\begin{equation}
\hat\sigma^{\rm 4F,coll}(\hat\tau)
=2\int_{\hat\tau}^1dz_1\int_{\frac{\hat\tau}{z_1}}^1dz_2\,
\left[\frac{\alpha_s}{2\pi}P_{qg}(z_1)L(z_1,\hat\tau)\right]
\left[\frac{\alpha_s}{2\pi}P_{qg}(z_2)L(z_2,\hat\tau)\right]
\hat\sigma^{\rm 5F}\left(\frac{\hat\tau}{z_1z_2}\right).
\label{parton4F}
\end{equation}
The physical interpretation of the result
Eq.~(\ref{parton4F}) is straightforward: in the limit of collinear emission,
the cross section for the parton process
(\ref{ggbbH}) is simply the $b\bar b\to H$ cross section convolved
with the probability that the incident gluons split in a $b\bar b$ pair.
This probability is logarithmically divergent as $m_b\to 0$,
and this is the origin of the two factors of $L(z_i,\hat\tau)$.

The arguments of the two collinear logarithms exhibit a dependence on the
momentum fractions $z_1,z_2$, Eq.~(\ref{logs}). This dependence is
subleading in the collinear limit $m_b\to 0$ and indeed it could be
neglected in this approximation; however, the class of subleading
terms induced by the factor $(1-z_i)^2/\hat\tau$ in Eq.~(\ref{logs}) is
of kinematical origin (it arises from the integration bounds on $t_1$
and $t_2$, as shown in Appendix~\ref{appxs}) and therefore universal
in some sense, as illustrated in Ref.~\cite{Maltoni:2012pa}.
We also note that the arguments of the two collinear logs depend on both
$z_1$ and $z_2$; this is to be expected, because the integration bounds
on $t_1$ and $t_2$ are related to each other.
However, in some cases (for example, if one
wants to relate the scale choice to a change of factorisation scheme, as in
ef.~\cite{Maltoni:2007tc}) 
a scale choice which only depends on the kinematics of 
each emitting line might be desirable. We have checked that the replacement
\begin{equation}
\log\left[\frac{M_H^2}{m_b^2}\frac{(1-z_i)^2}{z_1z_2}\right]
\to
\log\left[\frac{M_H^2}{m_b^2}\frac{(1-z_i)^2}{z_i}\right]
\label{rep}
\end{equation}
has a moderate effect on physical cross sections.
The replacement would make the scale at which the four- and five-flavor
scheme results are comparable lower by about 20/30\% but does not
qualtatively modify our arguments and results below.

The corresponding 4F scheme physical cross section in hadron
collisions at centre-of-mass energy $\sqrt{s}$ is given by
\begin{equation}
\sigma^{\rm 4F,coll}(\tau)=\int_\tau^1dx_1\,\int_{\frac{\tau}{x_1}}^1dx_2\,
g(x_1,\mu_F^2)g(x_2,\mu_F^2)
\hat\sigma^{\rm 4F,coll}\left(\frac{\tau}{x_1x_2}\right),
\end{equation}
where $g(x,\mu_F^2)$ is the gluon distributon function,
$\mu_F$ is the factorisation scale, and
\begin{equation}
\tau=\frac{M_H^2}{s}.
\end{equation}
After some (standard) manipulations, we get
\begin{align}
&\sigma^{\rm 4F,coll}(\tau)=
2\int_\tau^1dx_1\,\int_{\frac{\tau}{x_1}}^1dx_2\,
\hat\sigma^{\rm 5F}\left(\frac{\tau}{x_1x_2}\right)
\nonumber\\
&\quad
\int^1_{x_1}\frac{dz_1}{z_1}\,
\left[\frac{\alpha_s}{2\pi}
P_{qg}(z_1)L\left(z_1,z_1z_2\right)\right]
g\left(\frac{x_1}{z_1},\mu_F^2\right)
\int^1_{x_2}\frac{dz_2}{z_2}\,\left[\frac{\alpha_s}{2\pi}
P_{qg}(z_2)L\left(z_2,z_1z_2\right)\right]
g\left(\frac{x_2}{z_2},\mu_F^2\right).
\label{had4F}
\end{align}
We are now ready to assess the accuracy of the collinear approximation
in the 4F scheme. We first consider the total cross 
section. In table~\ref{table:xsec4F} we display
the total 4F scheme cross section for the production of a Higgs boson
at LHC 13 TeV for two values of the Higgs mass, namely
$M_H=125$ GeV and $M_H=400$ GeV. 
\begin{table}[htb]
\begin{center}
\begin{tabular}{|c||c|c|c|}
\hline
$M_H$ & exact & collinear ME & collinear ME and PS\\
\hline
\hline
125 GeV & 4.71 $\cdot\,10^{-1}$ pb & 5.15 $\cdot\,10^{-1}$ pb 
& 5.82 $\cdot\,10^{-1}$ pb \\
\hline
400 GeV & 5.42 $\cdot\,10^{-3}$ pb & 5.58 $\cdot\,10^{-3}$ pb 
& 5.91 $\cdot\,10^{-3}$ pb\\
\hline
\end{tabular}
\end{center}
\caption{\label{table:xsec4F}
Total cross sections for Higgs boson production at the LHC 13 TeV
in the 4F scheme. }
\end{table}
In the first column we give the exact leading order result;
the second column contains the cross section with the squared amplitude
approximated by its collinear limit, but the exact expression of the phase
space measure. Finally, in the third column we give the results
obtained with both the amplitude and the phase-space measure in the collinear
limit, which corresponds to the expression in Eq.~(\ref{had4F}).
From table~\ref{table:xsec4F} we conclude that the production of
large transverse momentum $b$ quarks, 
correctly taken into account in the 4F scheme, amounts to 
an effect of order 20\% on the total cross section
and tends to decrease with increasing Higgs mass.

\begin{figure}[ht]
\includegraphics[width=0.50\textwidth]{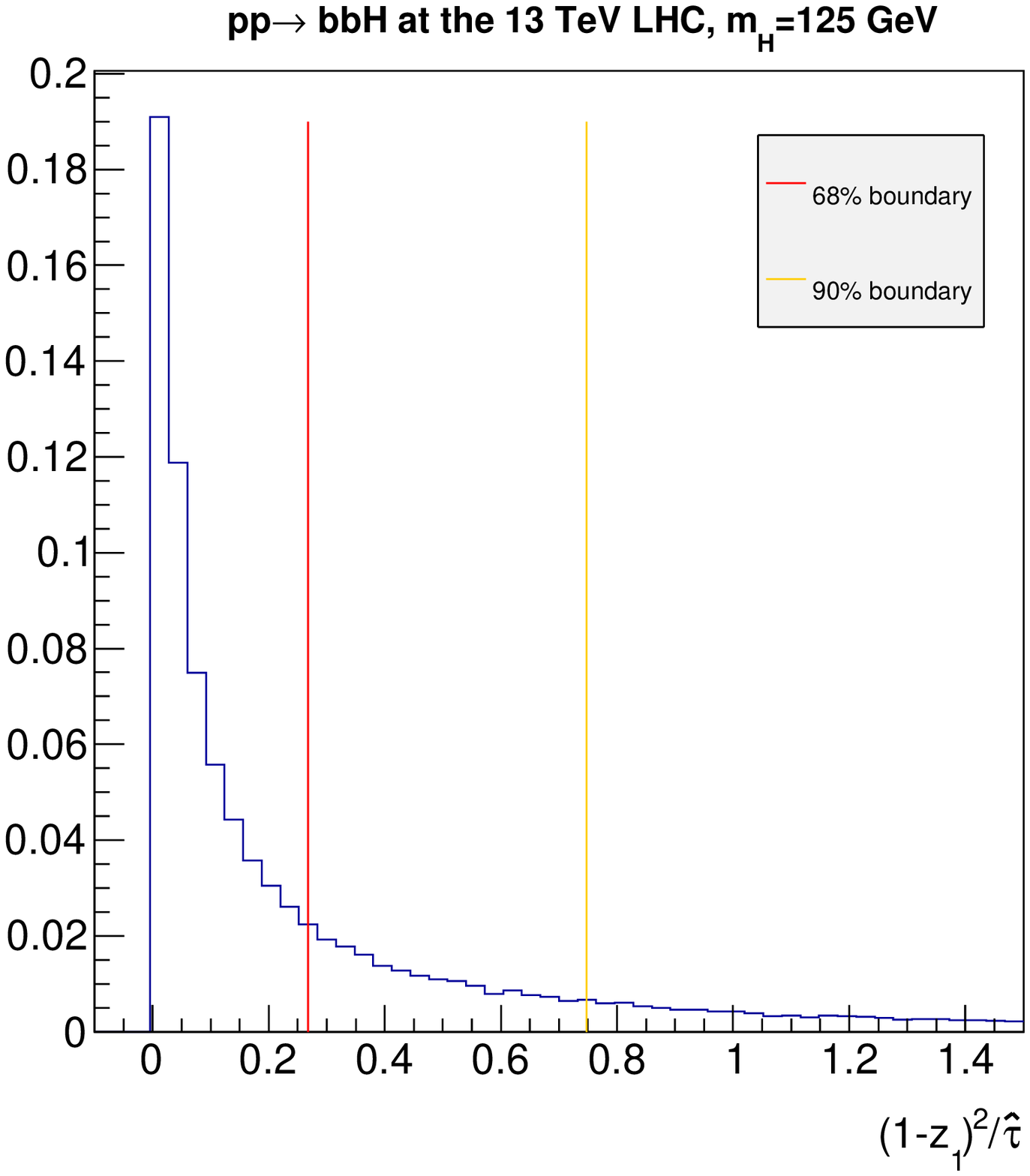}
\includegraphics[width=0.50\textwidth]{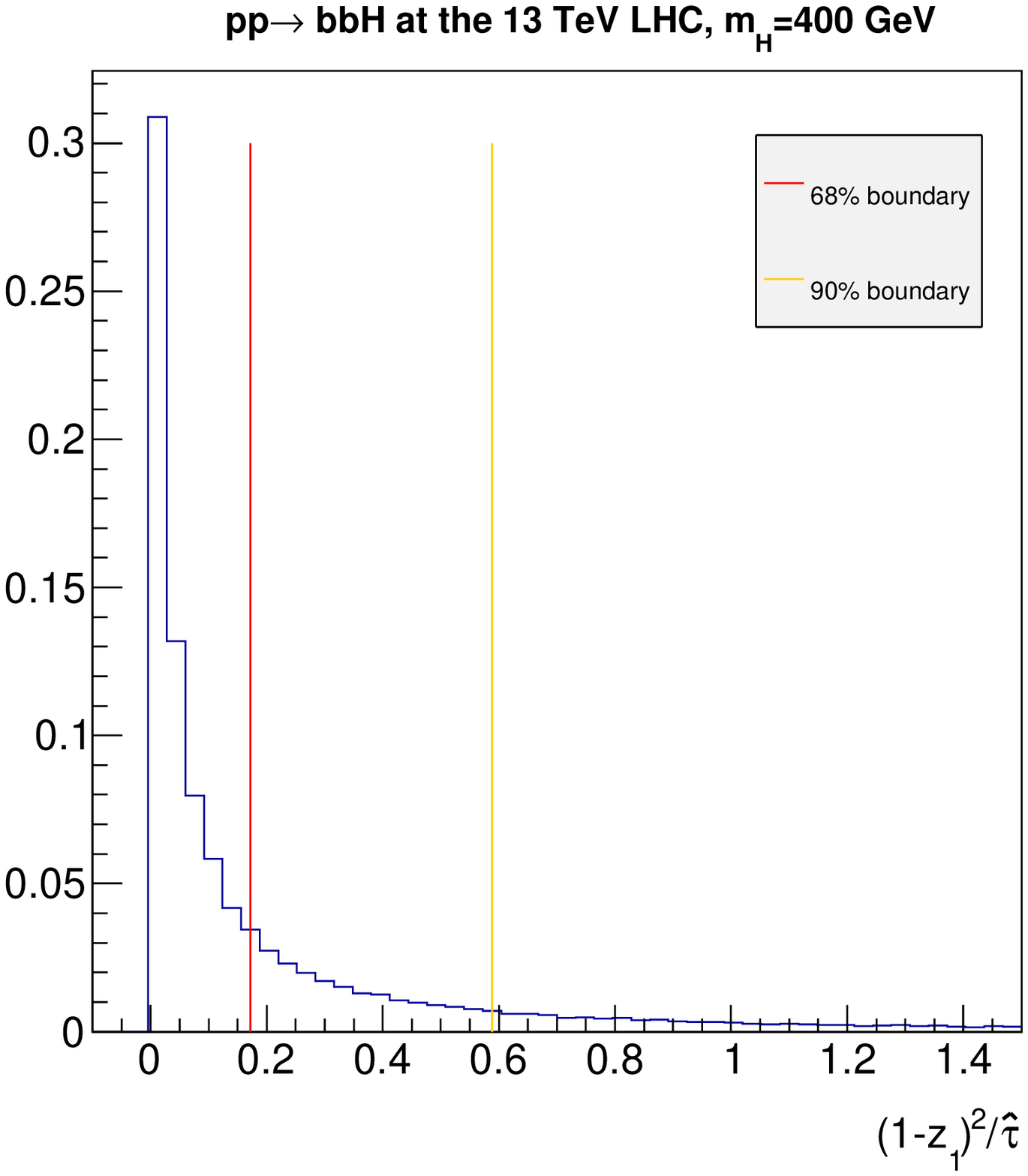}
\caption{\label{fig:zdist:lhc}
Normalised distribution (events/bin) of $(1-z_1)^2/\hat{\tau}$
for $b$-initiated Higgs production in
$pp$ collisions at LHC 13 TeV for $M_H=125$ GeV (left)
and $M_H=400$ GeV (right). 
Both $\mu_R$ and $\mu_F$ are set to $M_H$.
The vertical lines represent the values
below which 68\% and 90\% of events lie.}
\end{figure}

We now turn to an assessment of the numerical relevance of the
subleading terms included by the definition Eq.~(\ref{logs}) of
the collinear logarithms.
To this purpose we study the distribution of 
$(1-z_1)^2/(z_1z_2)$, which is the suppression
factor of $M_H^2/m_b^2$ in the arguments of the logs.
The results are displayed in Fig.~\ref{fig:zdist:lhc}
for Higgs production at the LHC at 13 TeV and for two different
values of the Higgs boson mass.
The two distributions behave in a similar way:
both are strongly peaked around values smaller than 1; in particular, the
68\% threshold is in both cases around $0.2$. This confirms that,
altough formally subleading with respect to $\log\frac{M_H^2}{m_b^2}$,
in practice the terms proportional to $\log\frac{(1-z_i)^2}{z_1z_2}$ 
give a sizeable contribution to the total cross section.

\begin{figure}[t]
\centering
\includegraphics[width=0.47\textwidth]{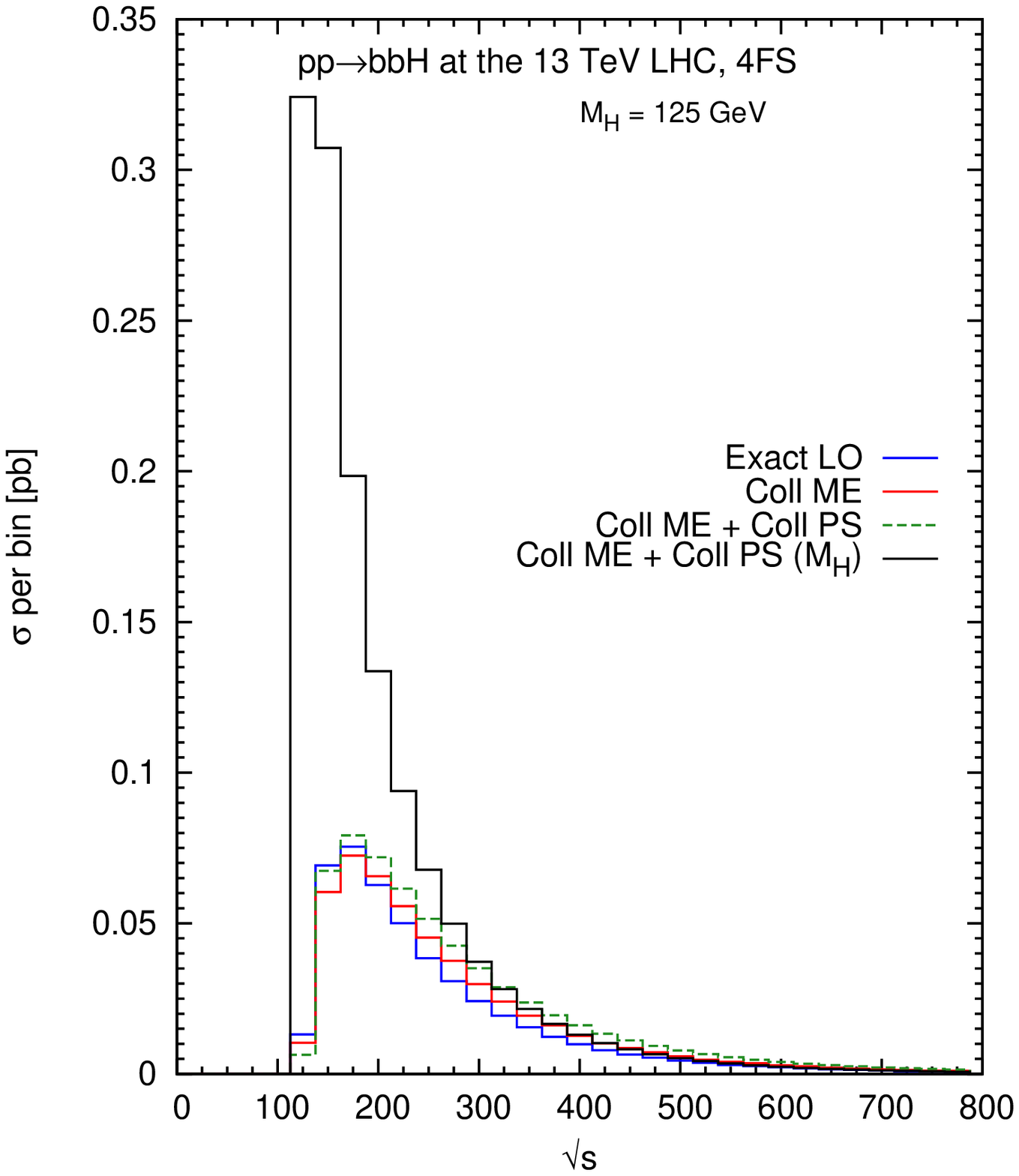}
\includegraphics[width=0.47\textwidth]{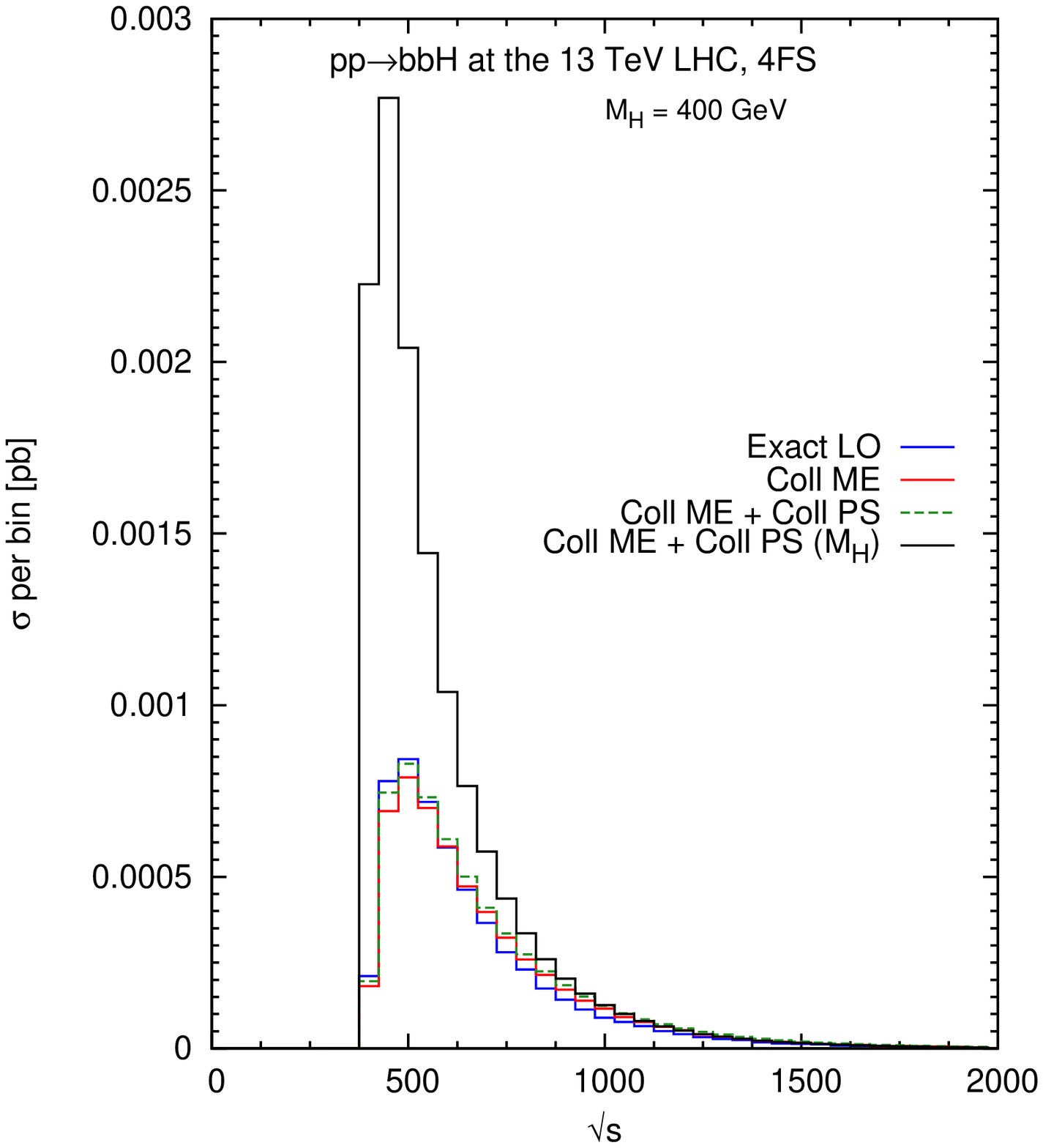}
\caption{\label{fig:sdist}
Distribution of the 4F scheme cross section as a function of the partonic 
centre-of-mass energy $\hat{s}$ for a Higgs of mass 125 GeV (above) 
and of mass 400 GeV (below). 
The solid line represents the full cross section
at leading-order, while the dashed line represents the collinear limit. }
\end{figure}

A further confirmation is provided by the distributions
in Fig.~\ref{fig:sdist}, where the full cross sections, together
with their collinear and double-collinear approximations, are plotted
as functions of the partonic centre-of-mass energy.
We see that the collinear cross section provides a good approximation to
the full 4F scheme result. In the same picture we show the collinear cross
section with the factors of $L(z_i,z_1z_2)$ replaced by
$\log\frac{M_H^2}{m_b^2}$ (solid black histogram). It is clear that in this case
the collinear cross section substantially differs from the exact result.

We now consider the 5F scheme, where the $b$ quark is treated as a
massless parton and collinear logarithms are resummed to all orders by
the perturbative evolution of the parton distribution function.
Eq.~(\ref{parton5F}) leads to a physical cross section
\begin{equation}
\sigma^{\rm 5F}(\tau) = 2\int_\tau^1dx_1\,b(x_1,\mu_F^2)
\int_{\frac{\tau}{x_1}}^1dx_2\,
b(x_2,\mu_F^2)\hat\sigma^{\rm 5F}\left(\frac{\tau}{x_1x_2}\right).
\label{had5Ffull}
\end{equation} 
In order to make contact with the 4F scheme calculation, we observe that
the $b$ quark PDF can be expanded to first order in $\alpha_s$:
\begin{equation} 
b(x,\mu_F^2)
 = \frac{\alpha_s}{2\pi}L_b
\int^1_x\frac{dy}{y}\, P_{qg}(y) g\left(\frac{x}{y},\mu_F^2\right)
+\order{\alpha_s^2}
=\tilde b^{(1)}(x,\mu_F^2)+\order{\alpha_s^2},
\label{btilde1}
\end{equation} 
where
\begin{equation}
L_b=\log\frac{\mu_F^2}{m_b^2}.
\end{equation}
Correspondingly, we may define a truncated 
5F cross section $\sigma^{\rm 5F,(1)}(\tau)$ which contains only
one power of $\log m_b^2$
for each colliding $b$ quark. This is obtained by
replacing Eq.~(\ref{btilde1}) in Eq.~(\ref{had5Ffull}) and performing
the same manipulations that led us to Eq.~(\ref{had4F}): we get
\begin{align}
\sigma^{\rm 5F,(1)}(\tau)&=2
\int_\tau^1dx_1\,\int_{\frac{\tau}{x_1}}^1
dx_2\,\hat\sigma^{5F}\left(\frac{\tau}{x_1x_2}\right)
\nonumber\\
&
\int^1_{x_1}\frac{dy}{y}\,\left[\frac{\alpha_s}{2\pi}
P_{qg}(y)L_b\right]g\left(\frac{x_1}{y},\mu_F^2\right)
\int^1_{x_2}\frac{dz}{z}\,\left[\frac{\alpha_s}{2\pi}
P_{qg}(z)L_b\right]g\left(\frac{x_2}{z},\mu_F^2\right).
\label{had5F}\end{align} 

Eq.~(\ref{had5F}) has exactly the same structure as the 4F scheme result in the
collinear approximation Eq.~(\ref{had4F}), except that the collinear 
logarithms have a constant argument. Hence, it corresponds to the
solid black curve in Fig.~\ref{fig:sdist}.
We are therefore led to suggest that the 5F scheme results be used
with a scale choice dictated by the above results,
similar to what we have illustrated in Ref.~\cite{Maltoni:2007tc}. 
Such a scale is defined so that the two schemes give
the same result:
\begin{equation}
\sigma^{\rm 5F,(1)}(\tau) 
=\sigma^{\rm 4F,coll}(\tau).
\end{equation}
The explicit expression of $\tilde\mu_F$ is simply obtained by equating
$\sigma^{\rm 5F,(1)}(\tau)$, Eq.~(\ref{had5F}), which is proportional to
$L_b^2=\log^2\frac{\mu_F^2}{m^2}$, and 
$\sigma^{\rm 4F,coll}(\tau)$, Eq.~(\ref{had4F}), and solving for 
$L_b^2$. The residual dependence on $\mu_F$ due to the gluon parton
density is suppressed by an extra power of $\alpha_s$ and can therefore be 
neglected; we adopt the standard choice $\mu_F=M$, with $M$ either the
Higgs mass or the $Z'$ mass.
The size of the logarithmic terms kept explicitly 
in the 4F case is determined by
arguments of the form $\frac{(1-z_i)^2}{\hat\tau}$. 
For $\sqrt{s}=13$ GeV, and $m_b=4.75$ GeV, 
we find the following values for $\tilde{\mu}_F$:
\begin{align}
b\bar b H, M_H=125\,{\rm GeV}:  \qquad \qquad & \tilde\mu_F \approx 0.36\,M_H
\nonumber\\
b\bar b Z', M_{Z'}=91.2\,{\rm GeV}:  \qquad \qquad & \tilde\mu_F 
\approx 0.38\,M_{Z'}
\nonumber\\
b\bar b Z', M_{Z'}=400\,{\rm GeV}:  \qquad \qquad & \tilde\mu_F 
\approx 0.29\,M_{Z'},
\end{align}
while for $\sqrt{s}=100$ TeV  and $m_t=173.1$ GeV,
we find
\begin{align}
t\bar t Z', M_{Z'}=1\,{\rm TeV}:  \qquad & \tilde\mu_F \approx 0.40\,M_{Z'}
\nonumber\\
t\bar t Z', M_{Z'}=5\,{\rm TeV}:  \qquad & \tilde\mu_F \approx 0.21\,M_{Z'}
\nonumber\\
t\bar t Z', M_{Z'}=10\,{\rm TeV}:  \qquad & \tilde\mu_F \approx 0.16\,M_{Z'}.
\end{align}
In both cases we have used the 
{\tt NNPDF30\_lo\_as\_0130} PDF set~\cite{Ball:2014uwa}, with the appropriate number of light 
flavors. We have explicitly checked that the choice of $\mu_F=M_H/4$
for the gluon PDF and for the strong coupling constant does not modify
in any significant way the value of $\tilde\mu_F$ that we
obtain. This is expected given that the gluon-gluon luminosity and the
dependence on $\alpha_s$ tend to compensate between numerator and
denominator.
We have also checked that,
after the replacement in Eq.~(\ref{rep}), the values of $\tilde\mu_F$
are typically about 20-30\% smaller.

We note that the scale $\tilde\mu_F$ is in general remarkably smaller
than the mass of the produced heavy particle.  As in the case of
single collinear logarithm, the reduction is more pronounced for
larger values of the mass of the heavy particle compared to the
available hadronic centre-of-mass energy.  The above results suggest
that a ``fair" comparison between calculations in the two schemes
should be performed at factorisation/renormalisation scales smaller
than the na\"ive choice $\mu_F=M_H$. This evidence backs up the
conclusions drawn in previous studies~\cite{Maltoni:2003pn}, although
perhaps with a slightly larger value in the case of Higgs boson,
$\tilde{\mu}\approx M_H/3$ rather than $M_H/4$.

The argument given above identifies a suitable choice for the
factorisation/renormalisation scales such that, at the Born level
and without resummation, the size of the logarithmic terms is
correctly matched in the two schemes. At this point, further
differences between the schemes can arise from the collinear
resummation as achieved in the 5F scheme and from mass (power-like)
terms which are present in the 4F scheme and not in the 5F one. Closely
following the arguments of Ref.~\cite{Maltoni:2003pn}, to which we
refer the interested reader for more details, we now numerically
quantify the effect of the resummation. A careful study of the impact
of power-like terms can be found in
Refs.~\cite{Forte:2015hba,Bonvini:2015pxa,Bonvini:2016fgf}. These terms
have been found to have an impact no stronger than a few percents.

Starting from Eq.~(\ref{btilde1}), one can assess the accuracy of the
$\order{\alpha_s^1}$ ($\order{\alpha_s^2}$) approximations compared to
the full $b(x,\mu^2)$ resummed expression. The expansion truncated at
order $\alpha_s^p$, often referred to as $\tilde b^{(p)}(x,\mu^2)$ in
the literature, does not feature the full resummation of collinear
logarithms, but rather it contains powers $n$ of the collinear log
with $1\le n\le p$.

\begin{figure}[htb] \centering
\includegraphics[width=0.45\textwidth]{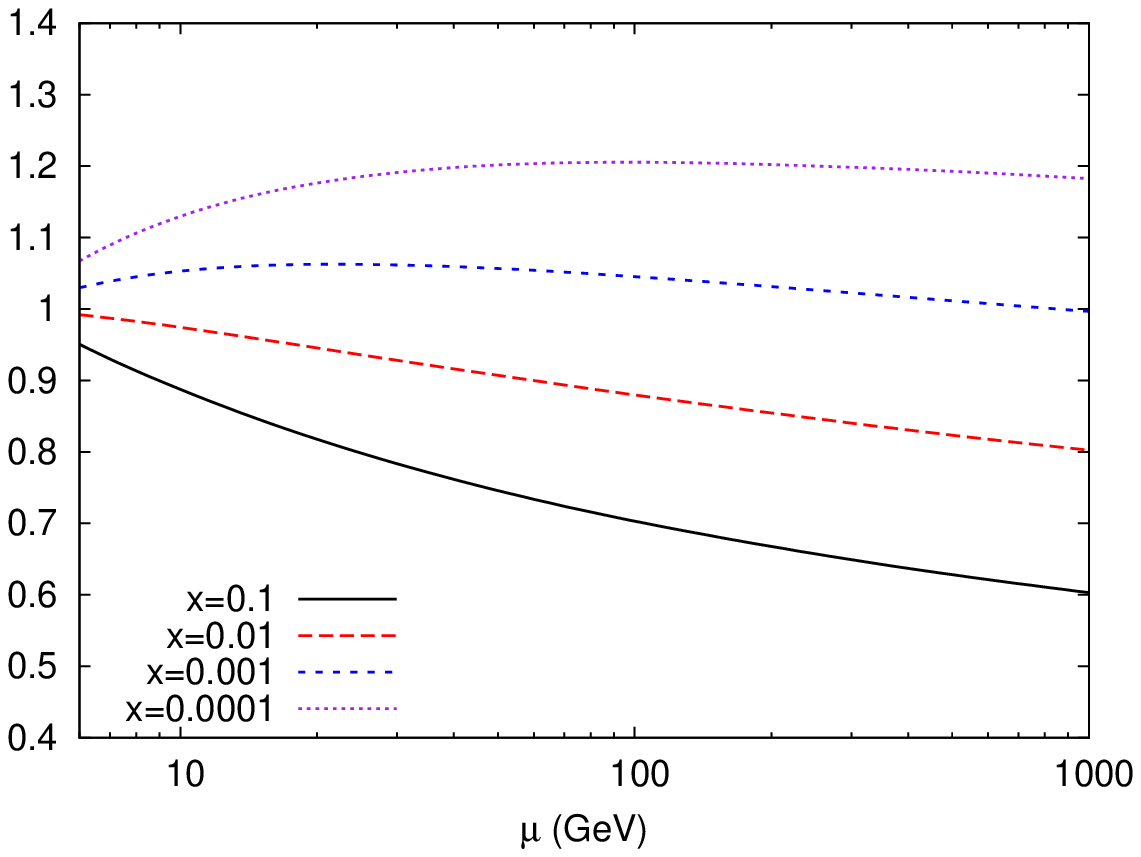} 
\includegraphics[width=0.45\textwidth]{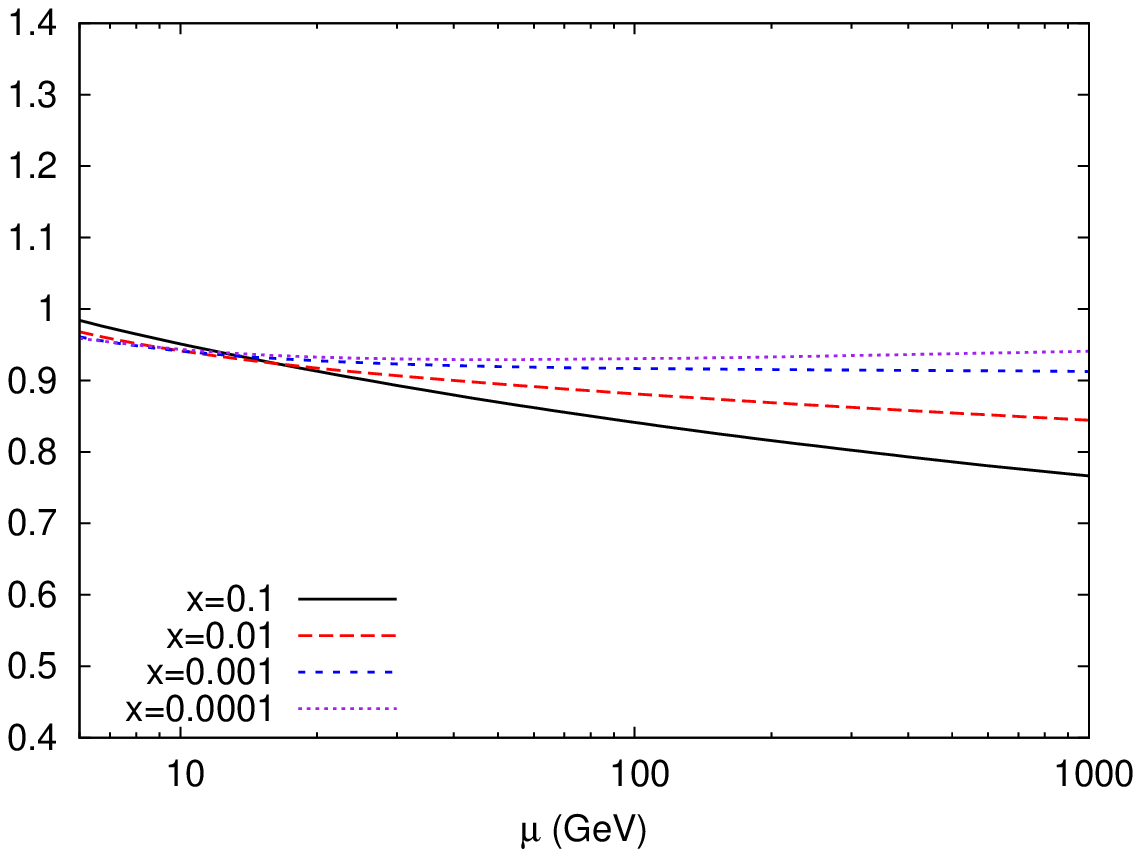} 
\caption{\label{fig:b-tilde} The ratio $\tilde b^{(p)}/b$ for $p=1$ (left)
  and $p=2$ (right) as a function of the scale $\mu$ for for different
  values of $x$.  The $n_f=4$ and $n_f=5$ sets of the {\tt NNPDF3.0}
  family (with $\alpha_s(M_Z)=0.118$) are associated to the
  $\tilde{b}$ and $b$ computations respectively.}
\end{figure}

In Fig.~\ref{fig:b-tilde} we display  the ratio 
$\frac{\tilde b^{(p)}(x,\mu^2)}{b(x,\mu^2)}$ for $p=1,2$ (using the same set
of PDFs adopted throughout this work) as a function of 
the scale $\mu^2$ for various values of the momentum fraction $x$.
Deviations from one of these curves are an indication of the size of
terms of order $\order{\alpha_s^{p+1}}$ and higher,
which are resummed in the QCD evolution of the bottom quark PDFs.  
As observed in our previous work, at LO higher-order logarithms are
important and $\tilde{b}^{(1)}(x,\mu^2)$ is a poor approximation of
the fully resummed distribution function.  In particular, it
overestimates the leading-log evolution of the $b$ PDF by 20\% at very
small $x$ and it underestimates it up to 30\% at intermediate values
of $x$.  On the other hand, at NLO the explicit collinear logs present
in a NLO 4F scheme calculation provide a rather accurate approximation
of the whole resummed result at NLL; significant effects, of order up
to 20\%, appear predominantly at large values of $x$. 

\begin{figure}[ht] \centering
\includegraphics[width=0.45\textwidth]{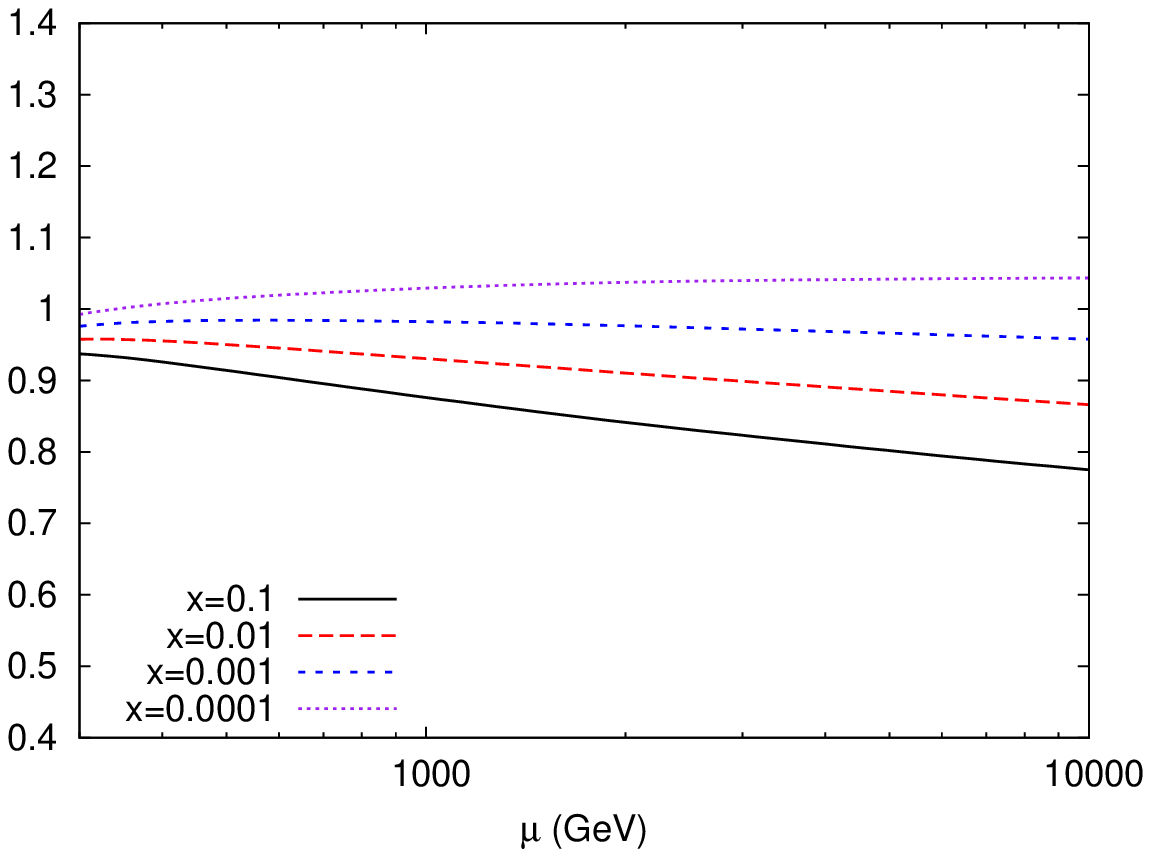} 
\includegraphics[width=0.45\textwidth]{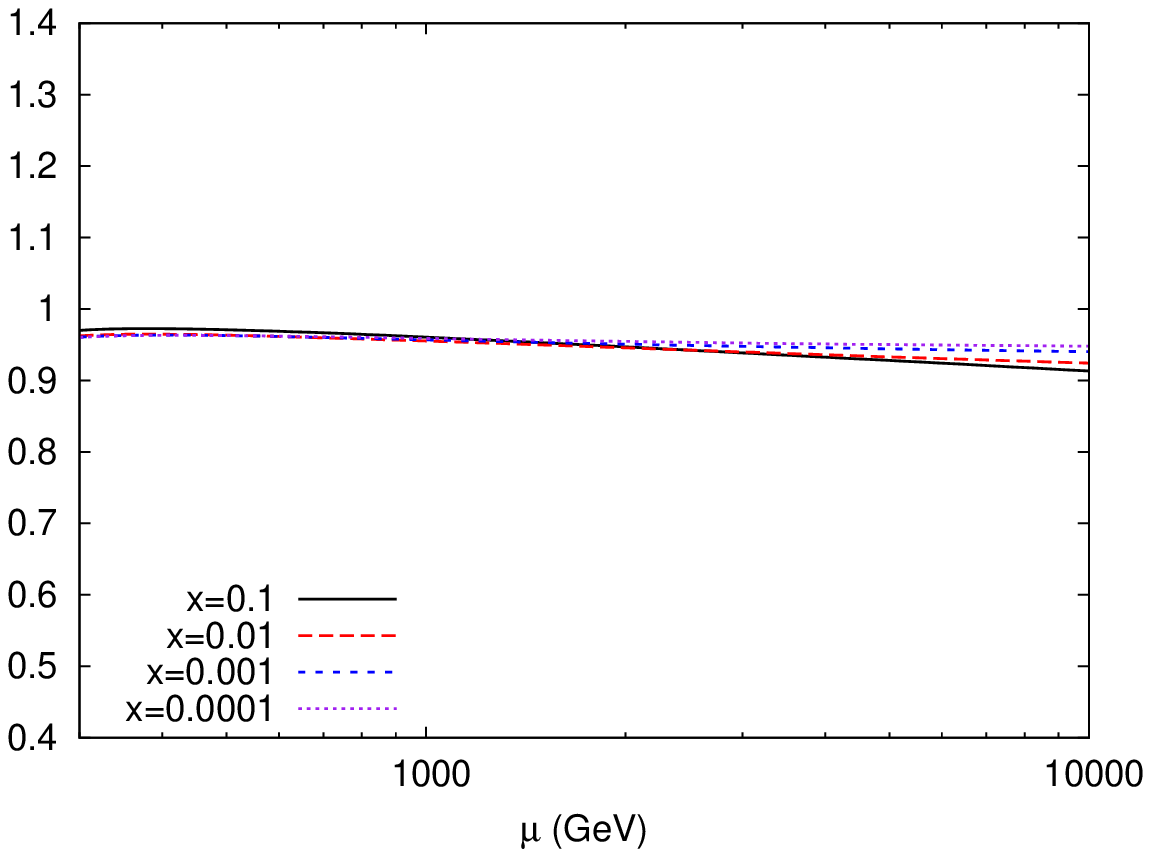} 
\caption{\label{fig:t-tilde} Ratio $\tilde{t}/t$ at LO (left) and NLO
  (right) for several values of $x$ as a function of the scale $\mu$.
  The $n_f=5$ and $n_f=6$ sets of the {\tt NNPDF3.0} family (with
  $\alpha_s(M_Z)=0.118$) are associated to the $\tilde{t}$ and $t$
  computations respectively.}
\end{figure}

A similar behaviour characterises the top-quark PDFs. In Fig.~\ref{fig:t-tilde} 
the ratio between the truncated top-quark PDFs $\tilde{t}$  and 
the evolved PDFs $t(x,\mu^2)$ is displayed for four different values of $x$ 
and varying the factorization scale $\mu$.
We see that for the top-quark PDF at NLO, the difference between the 2-loop
approximated PDF $\tilde{t}^{(2)}(x,\mu^2)$ and the fully evolved PDF
$t(x,\mu^2)$ is very small (of the order of $5$\%) unless very
high scales and large $x$ are involved. A comparable behaviour was
observed in Ref.~\cite{Dawson:2014pea}.

\section{Different heavy quark schemes: numerical results}
\label{sec:numeric}
In this Section, we consider the production of Higgs and neutral vector bosons
via $b\bar b$ fusion at the LHC and the production of heavy vector bosons
in $t\bar t$ collisions at a future high energy hadron collider.
We compare predictions for total rates obtained at the
highest available perturbative order in the 4F and 5F schemes 
at the LHC and in the 5F and 6F schemes at a future 100 TeV collider.

\subsection{LHC Run II}

\subsubsection{Bottom-fusion initiated Higgs production}

Although in the SM the fully-inclusive $b\bar{b}\to H$ cross section
is much smaller than the other Higgs production channels (gluon
fusion, vector boson fusion, $W$ and $Z$ associated Higgs production)
and its rate further decreases when acceptance cuts on the associated
$b$ quarks are imposed, this production process can be important in
several non-standard scenarios.  For example, in supersymmetric
models Higgs production in association with $b$ quarks can become a
dominant production channel when couplings are enhanced with respect
to the Standard Model.  More specifically, in models featuring a
second Higgs doublet the rate is typically increased by a factor
$1/\cos^2\beta$ or $\tan^2\beta$, with $\beta = v_1/v_2$ being the
ratio of two Higgs vacuum expectation values.

Calculations for $b$-initiated Higgs productions have been made
available by several groups.  The total cross section for this process
is currently known up to next-to-next-to-leading order (NNLO) in the
5F scheme~\cite{Harlander:2003ai} and up to next-to-leading order
(NLO) in the 4F scheme~\cite{Dittmaier:2003ej,Dawson:2005vi}.  Total
cross section predictions have been also obtained via matching
procedures that include the resummation of the collinear logarithms on
one side and the mass effects on the other, without double counting
common terms.  A first heuristic proposal, which has been adopted for
some time by the HXSWG LHC, is based on the so-called Santander
matching~\cite{Harlander:2011aa} where an interpolation between
results in the 4F and in the 5F schemes is obtained by means of a
weighted average of the two results.  Several groups have provided
properly matched calculations based on a thorough quantum field theory
analysis, at NLO+NLL and beyond via the FONLL
method~\cite{Forte:2015hba} and an effective field theory
approach~\cite{Bonvini:2015pxa,Bonvini:2016fgf} that yield very
similar results.

Fully differential calculations in the 4F scheme up to NLO(+PS)
accuracy have been recently made available~\cite{Wiesemann:2014ioa} in
\amc\ ~\cite{Alwall:2014hca} and work is in progress in the {\sc
  SHERPA}~ framework~\cite{Gleisberg:2008ta}. These studies conclude
that the 4F scheme results, thanks to the matching to parton showers,
are generally more accurate than the pure 5F scheme counterparts,
especially for observables which are exclusive in the $b$-quark
kinematics.  On the other hand, for inclusive observables the
differences between 4F and 5F schemes are mild if judicious choices
for scales are made.  The assessment of the size of such effects and
their relevance for phenomenology is the purpose of this section.

\begin{figure}[ht]
\centering
\includegraphics[width=0.45\textwidth]{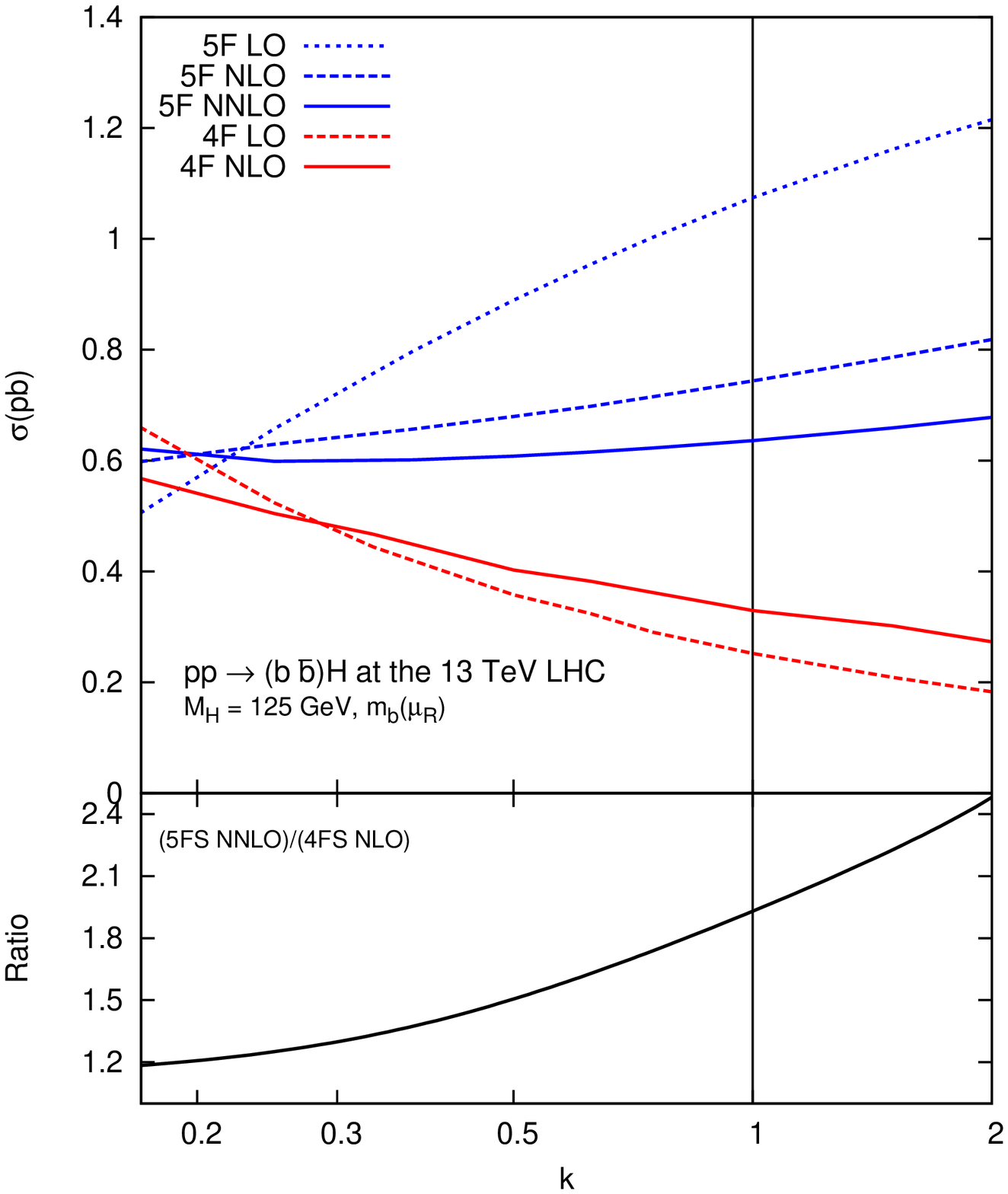}
\includegraphics[width=0.45\textwidth]{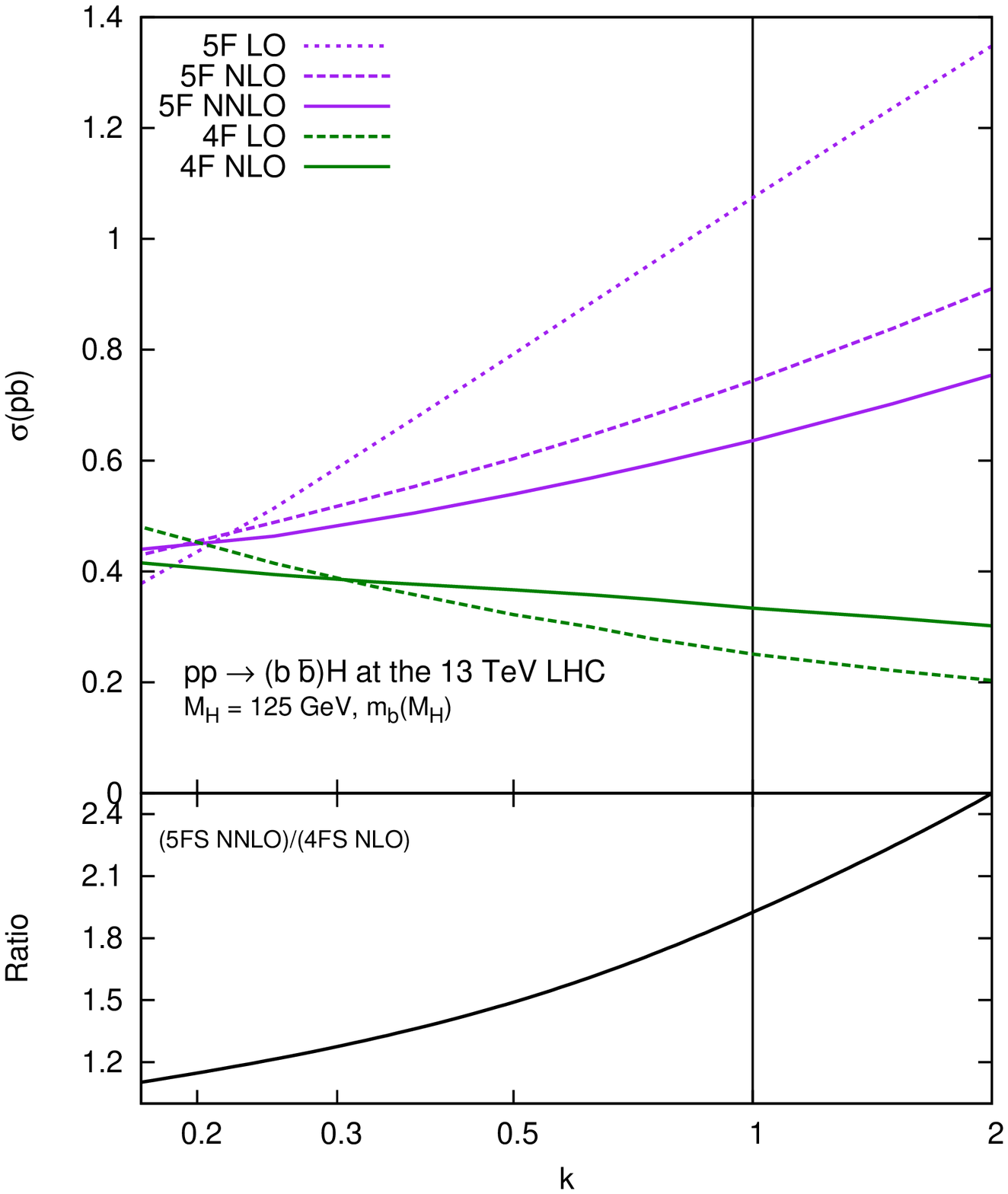}
\caption{\label{fig:bbH} Cross sections for the production of the SM
  Higgs boson via $b\bar b$ fusion ($y_b^2$ term only) in the 5F and
  4F schemes for LHC 13 TeV as functions of $k=\mu/M_H$, with
  $\mu_F=\mu_R=\mu$. Terms proportional to $y_by_t$ in the NLO 4F
  scheme have been neglected. Results with the running $b$ mass
  computed at a fixed scale $M_H$ are also shown (right plot).  In the
  inset the ratio between the 5F NNLO prediction and the 4F scheme NLO
  prediction is displayed.}
\end{figure}

\begin{figure}[ht]
\centering
\includegraphics[width=0.45\textwidth]{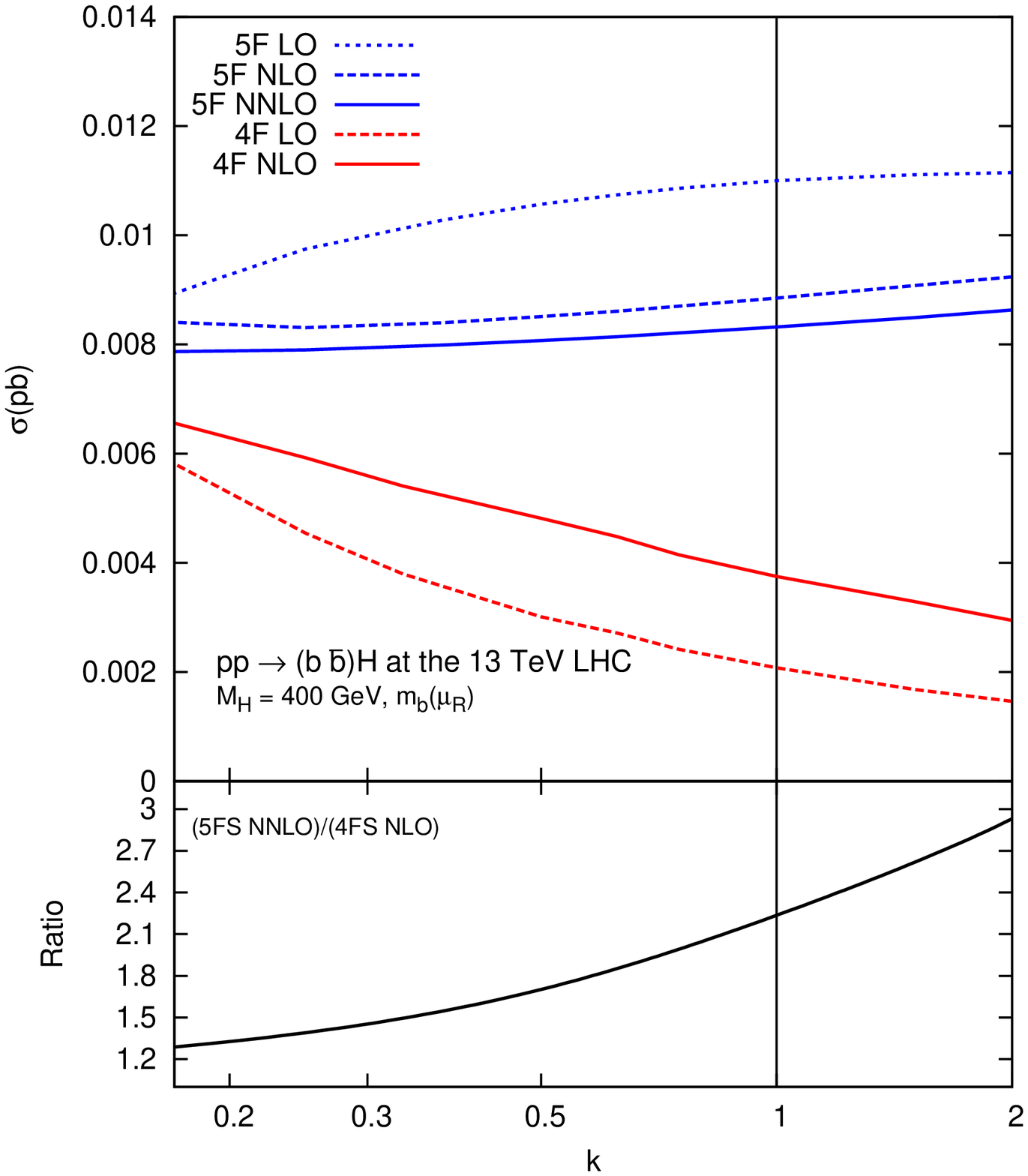}
\includegraphics[width=0.45\textwidth]{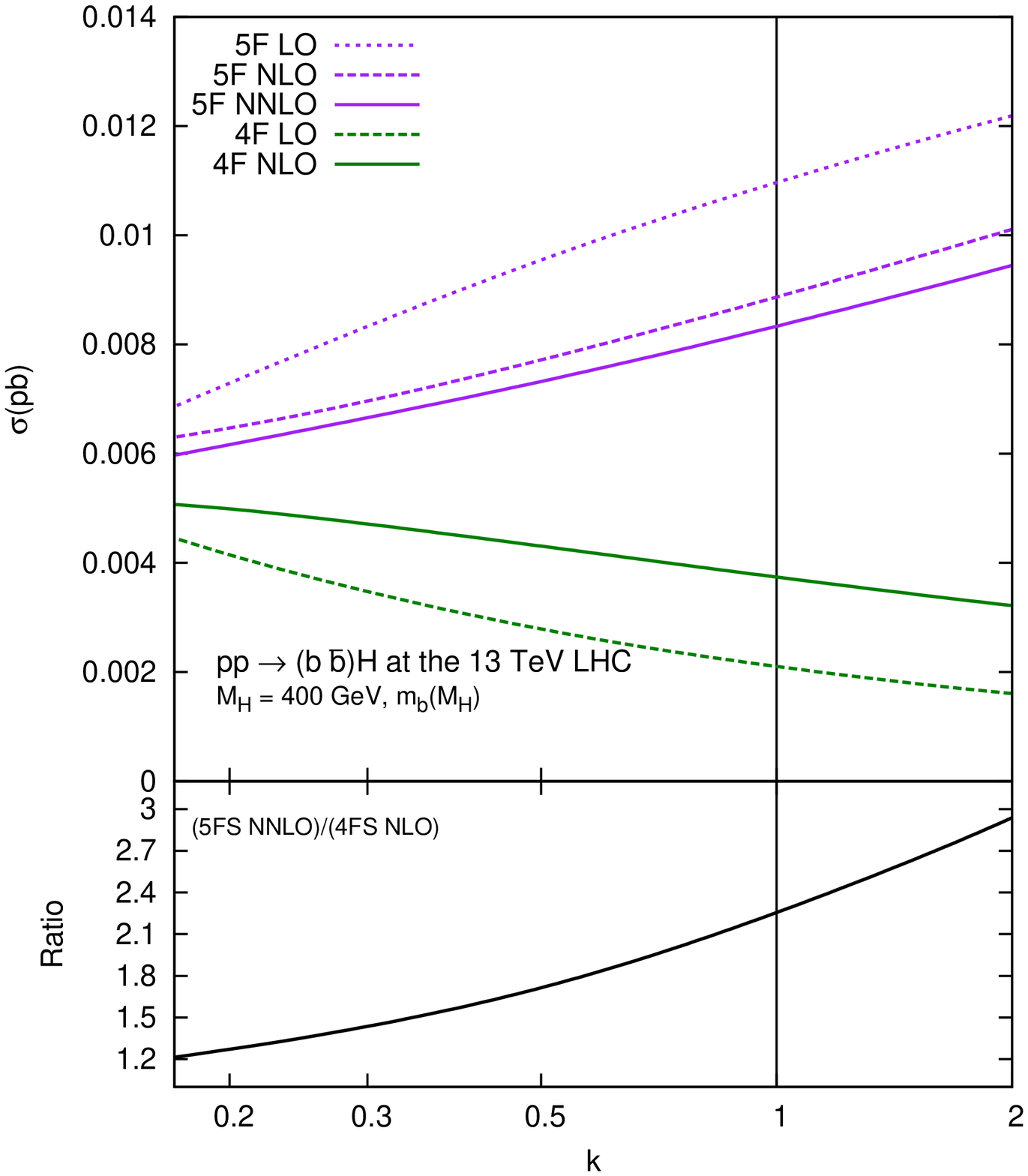}
\caption{\label{fig:bbHheavy}
Same as Fig.~\ref{fig:bbH} with $M_H=$ 400 GeV}
\end{figure}

We first compare the size and the scale dependence of the 4F and 5F
scheme predictions from leading-order up to the highest available
perturbative order, namely NLO in the case of the 4F scheme and
NNLO in the case of the 5F scheme cross sections.
Results are shown in Figs.~\ref{fig:bbH} and~\ref{fig:bbHheavy} for
the SM Higgs ($M_H=125$ GeV) and a heavier Higgs ($M_H=400$ GeV) respectively. 
The 4F scheme cross section has been generated using the public version of
\amc~\cite{Alwall:2014hca}.
In the case of the 5F scheme calculation, the cross section has been
computed with {\sc SusHi}~\cite{Harlander20131605} and the LO and NLO
results have been cross-checked against the output of \amc.  
The input PDFs belong to the {\sc NNPDF3.0} family
\cite{Ball:2014uwa} and the $n_f=4$ set was used in association with
the 4F scheme calculation, while the $n_f=5$ set was associated with
the 5F scheme calculation, consistently with the
perturbative order of the calculation, and with $\alpha_s^{\rm 5F}(M_Z)=0.118$.  
Both the renormalisation and factorisation scales have been taken to be equal to
$kM_H$, with $0.15 \le k \le 2$.

The treatment of the Higgs Yukawa
coupling to $b$ quarks deserves some attention. Different settings may
cause large shifts in theoretical predictions. Here we use the
$\msbar$ scheme; the running $b$ Yukawa $y_b(\mu)$ 
is computed at the scale $\mu_R$ (left plots). We have checked
that computing the Yukawa at the fixed value of $M_H$
does not modify our conclusions (right plots).
The numerical value of $m_b(\mu_R)$ is obtained from $m_b(m_b)$ 
by evolving up to $\mu_R$ at 1-loop (LO), 2-loops (NLO) or 3-loops
(NNLO) with $n_f=4$ or $n_f=5$, depending on the scheme.
The numerical value of $m_b(m_b)$ is taken to be equal to the pole
mass $m_b^{\rm pole} = 4.75$~GeV at LO (in both the 4F and 5F schemes),
$m_b(m_b) = 4.16$ GeV at NLO in the 5F scheme and $m_b(m_b) = 4.34$
GeV in the 4F scheme (consistently with the settings adopted in 
Ref.~\cite{Wiesemann:2014ioa})
and finally $m_b(m_b) = 4.18$~ GeV at NNLO in the 5F scheme,
consistently with the latest recommendation of the Higgs cross section
working group\footnote{The pole mass value that we use in our
  calculation is slightly different from the latest recommendation
  $m_b^{\rm pole} = 4.92$~GeV as well as from the value used in the
  PDF set adopted in our calculation $m_b^{\rm pole} = 4.18$~GeV,
  however our results are not sensitive to these small variations
  about the current central value.}. 

The 4F and 5F scheme curves at leading order show an opposite
behaviour: in the 4F scheme the scale dependence is driven by the
running of $\alpha_s$ and therefore decreases with the scale, while
the 5F scheme case it is determined by the scale dependence of the
$b$-quark PDF which in turn leads to an increase.  The inclusion of
higher orders in both calculations drastically reduces the
differences; nonetheless, it is clear from Figs.~\ref{fig:bbH}
and~\ref{fig:bbHheavy} that around the central scale $k=1$ the best 5F
scheme prediction exceeds the highest order 4F scheme prediction by a
large amount, about 80\%.  We also observe that 4F and 5F scheme
predictions are closer at lower values of the scale.  The scale
dependence of the 4F scheme NLO calculation is approximately of the
same size as that of the 5F scheme NLO calculation, while it is
stronger than the scale dependence of the 5F scheme NNLO calculation,
as expected, since in the latter the collinear logarithms are
resummed.

In Fig.~\ref{fig:bbHheavy} the same curves are displayed for a heavier
Higgs, $M_H=400$ GeV. As observed in Ref.~\cite{Maltoni:2012pa}, for
heavier final state particles differences between schemes are
enhanced. In particular, at the central scale the NNLO 5F scheme
prediction exceeds the 4F scheme case by a factor of two. Also in this case, at smaller
values of the scale the difference is significantly reduced.

This behaviour corresponds to that expected from our analysis
presented in sect.~\ref{sec:analytic}.  Comparing calculations at
${\tilde\mu}_F=0.36\, M_H$ for $M_H=125$ GeV and ${\tilde\mu}_F=0.29\,
M_H$ for $M_H=400$ GeV, the differences between the predictions in the
4F and 5F scheme reduce to about 30-35\%, a difference that can be
accounted for by considering first the (positive) effects of
resummation included in the 5F scheme calculation with respect to the 4F one
and second the power-like quark-mass corrections that are not included
the 5F calculation and estimated to be around $-$2-5\%, see
Ref.~\cite{Forte:2015hba,Bonvini:2015pxa,Bonvini:2016fgf}.

\begin{figure}[ht] \centering
\includegraphics[width=0.5\textwidth]{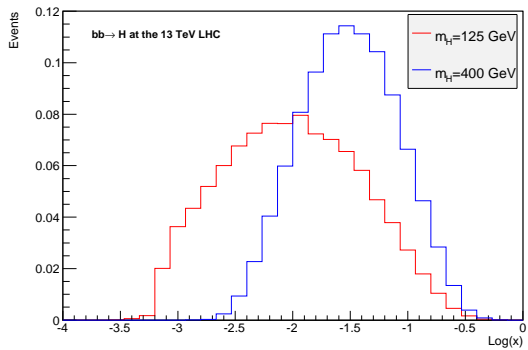} 
\caption{\label{fig:bbh-xb} Normalised distribution of the momentum fraction
  $x$ carried by the $b$ quark in $b\bar b$ initiated
  Higgs production, in the 5F scheme at leading order for LHC 13
  TeV, for $M_H=125$ GeV (red curve) and $M_H = 400$ GeV (blue
  curve).}
\end{figure}

The effects of the resummation are easy to quantify by establishing
the range of $x$ which gives the dominant contribution
to Higgs production via $b\bar b$ collisions. To this purpose,
we show in Fig.~\ref{fig:bbh-xb} the $x$ distribution in the 
leading-order bottom-quark fusion Higgs production in the 5F scheme.
We observe that the $x$ distribution has its maximum around
$x\approx 10^{-2}$ for the Standard Model Higgs;
for such values of $x$, the resummation of collinear logarithms
is sizeable: the difference between the fully resummed $b$ PDF 
and $\tilde b^{(2)}$ becomes as large as 10 to 15\% for scales between 100 and
400 GeV.  Note that we expect twice
the effect of a single $b$ quark in the case of processes
with two $b$ quarks in the initial state, which amounts to a
difference of 20-25\% from resummed logarithms at
$\order{\alpha_s^3}$ and higher between the collinear
approximation of the 4F scheme calculation and the 5F scheme calculation.

\begin{figure}[ht] \centering
\includegraphics[width=0.45\textwidth]{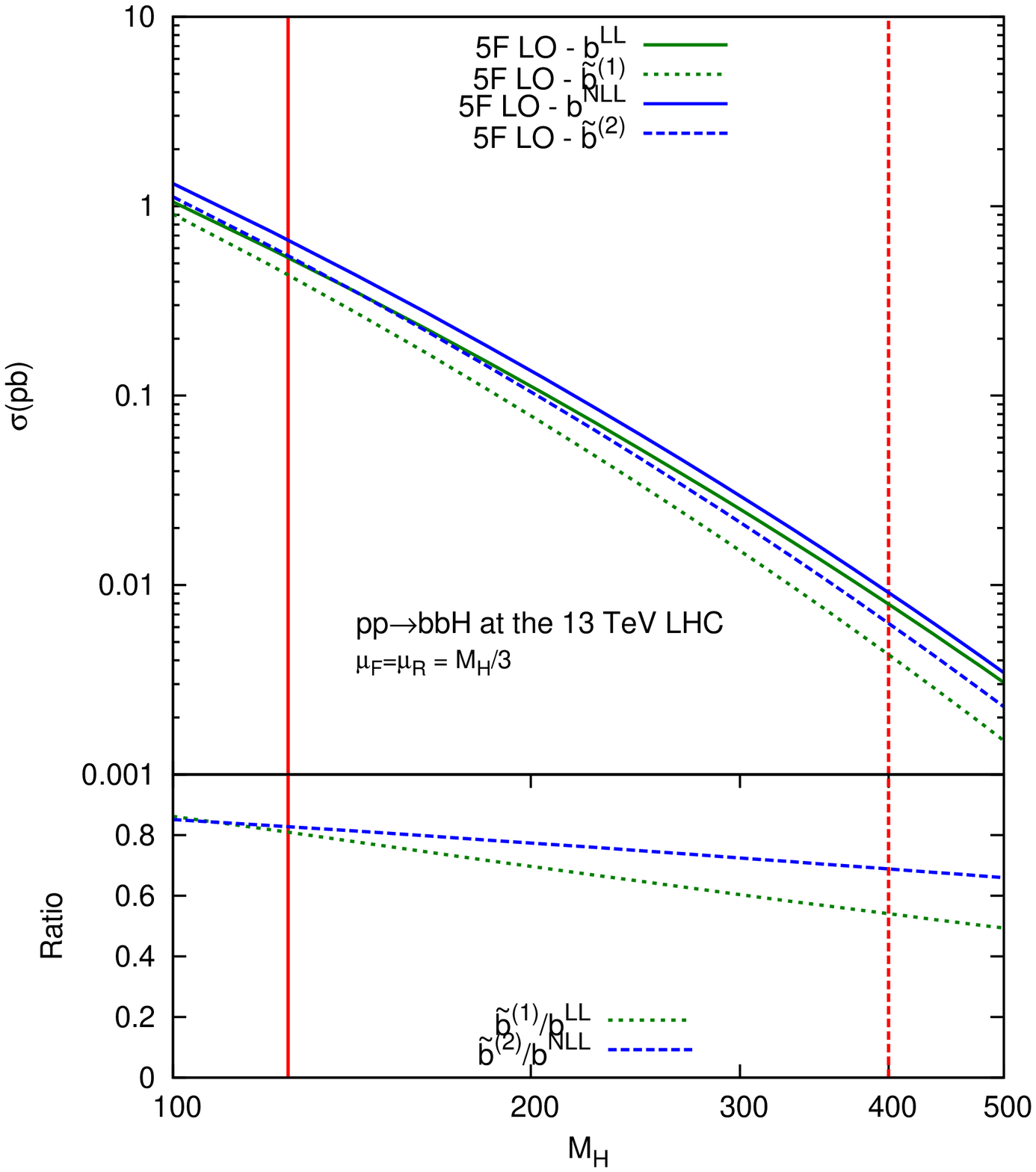} 
\includegraphics[width=0.45\textwidth]{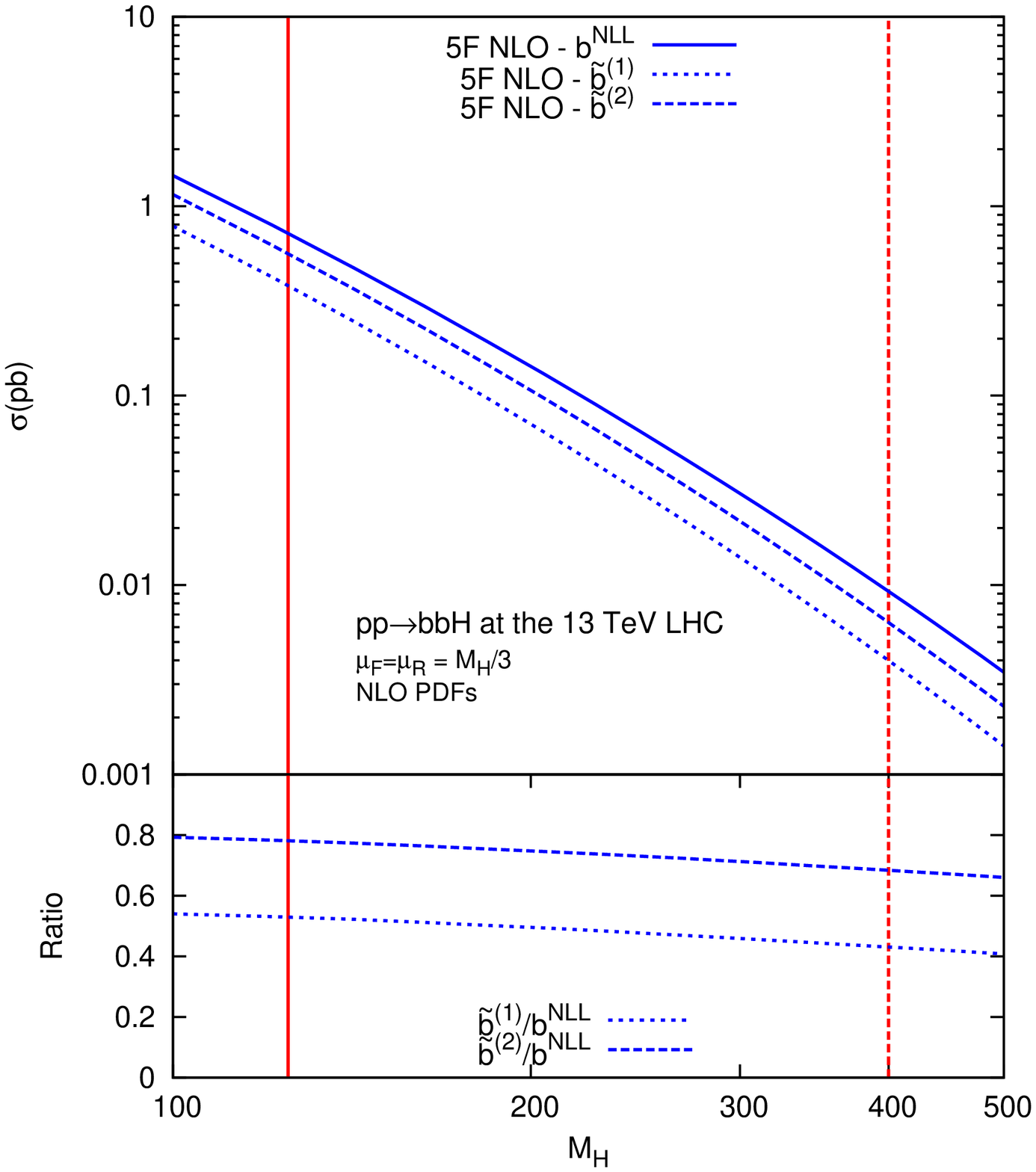} 
\caption{\label{fig:btilde-ratio-mh3}
Higgs production cross section
via $b\bar b$ fusion at LO (left) and NLO (right) as a function of $M_H$,
computed either with the fully resummed $b$ quark PDF at LL or NLL,
or with the truncated PDF $\tilde b^{(p)}$ with $p=1,2$,
with $\mu=\mu_F=\mu_R=M_H/3$.}
\end{figure}

This expectation is confirmed by the 
curves in Fig.~\ref{fig:btilde-ratio-mh3}, where we plot the
5F scheme cross section at LO (left panel) and NLO (right panel) as a function
of the Higgs mass in the range 100
GeV to 500 GeV, with $\mu_R=\mu_F=M_H/3$.  
The cross sections are computed with the same settings as in
Fig.~\ref{fig:bbH}. In the same panel we present the cross sections
with the $b$ PDF replaced by the $\tilde{b}^{(p)}$ truncated PDF
computed at order $p=1$ and at order $p=2$, together with the relevant
ratios.  We observe that, for a sensible value of the factorisation
and renormalisation scales, as per the one suggested in this paper
$\tilde{\mu}_F\sim M_H/3$, the effect of neglecting the higher order
logs resummed in the $b$ PDF evolution beyond the ones included in the
second order expansion of the $b$ PDF, $\tilde{b}^{(2)}$, is smaller
than 20\% for the SM Higgs mass and of about 30\% for a heavier
Higgs. Similar conclusions are drawn if the NLO cross section is
considered instead, as in the right hand-side panel.  If instead we
had taken as the central scale choice $\mu_R=\mu_F=M_H$ the effects of
the resummation of higher order logs would appear much more
significant.

\begin{figure}[ht] \centering
\includegraphics[width=0.45\textwidth]{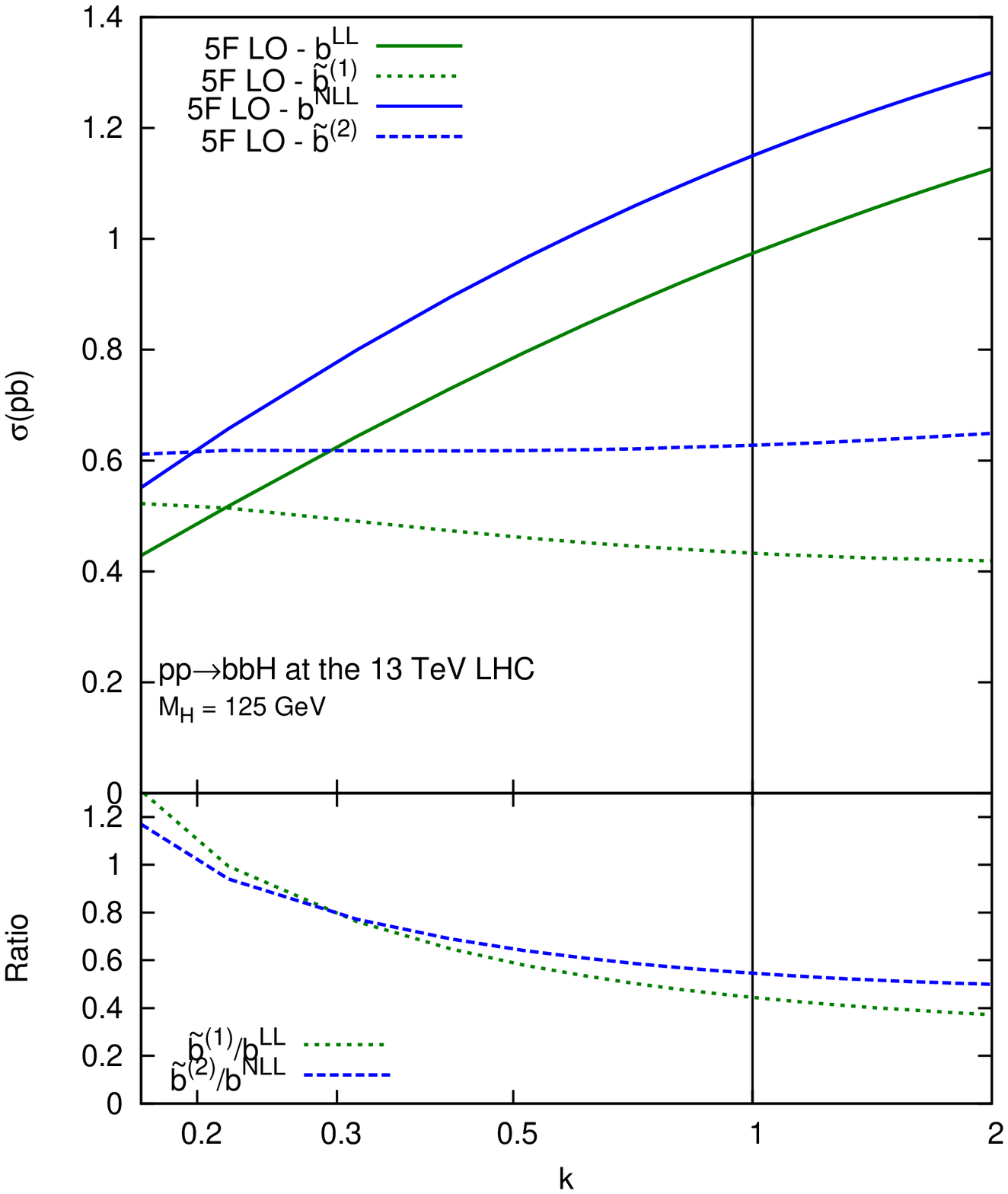} 
\includegraphics[width=0.45\textwidth]{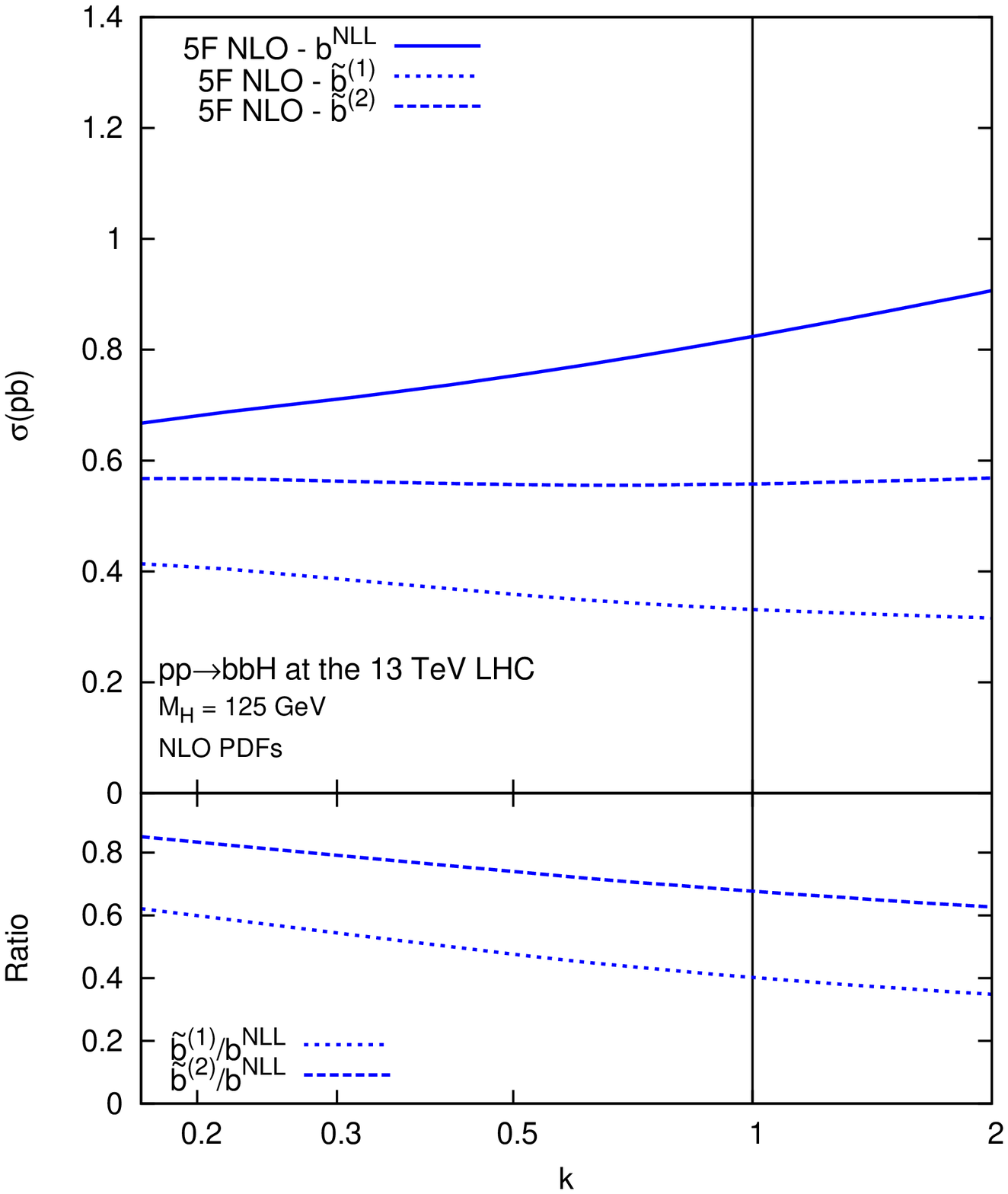} 
\caption{\label{fig:btilde-ratio-2}
Standard Model Higgs production cross section
via $b\bar b$ fusion at LO (left) and NLO (right) as a function of $k=\mu/M_H$,
with $\mu=\mu_R=\mu_F$,
computed either with the fully resummed $b$ quark PDF at LL or NLL,
or with the truncated PDF $\tilde b^{(p)}$ with $p=1,2$.}
\end{figure}

The scale dependence of the Standard Model Higgs cross section
is studied in Fig.~\ref{fig:btilde-ratio-2}. The plots confirm the
findings that the assessment of the effect of the higher-order logs
resummed in a 5F scheme calculation strongly depends on the scale at
which the process is computed and that at a scale close to $\tilde{\mu}_F$
the effects of higher order logs are quite moderate, while they become
significant if the naive hard scle of the process is chosen.

\subsubsection{Bottom-fusion initiated $Z'$ production}

A similar analysis can be carried out for the case of $Z$ production.
$Z$-boson production in association with one or two $b$-jets has a
very rich phenomenology. It is interesting as a testbed of our
understanding of QCD and it enters in precision measurements
(Drell-Yan at the LHC or indirectly in the $W$ mass determination). In
addition, it represents a crucial irreducible background for several
Higgs production channels at the LHC.  For the SM Higgs boson,
$Zb\bar{b}$ production is a background to $ZH$ associated production
followed by the decay of the Higgs into a bottom-quark pair.  Finally,
this process is a background to searches for Higgs bosons with
enhanced $Hb\bar b$ Yukawa coupling.

Calculations for bottom-initiated $Z$ production have been made
available by several groups. The $Zb\bar{b}$ production cross section
was originally computed (neglecting the $b$ quark mass) in
Ref.~\cite{Campbell:2000bg} for exclusive 2-jet final states.  The
effect of a non-zero $b$ quark mass was considered in later
works~\cite{FebresCordero:2008ci,Cordero:2009kv} where the total cross
section was also given. More recently, in Ref.~\cite{Frederix:2011qg}
leptonic decays of the $Z$ boson have taken into account, together
with the full correlation of the final state leptons and the parton
shower and hadronisation effects. The total cross section for
$Zb\bar{b}$ in the 5F scheme has been computed at NNLO accuracy for
the first time in Ref.~\cite{Maltoni:2005wd}.
 
Bottom-initiated $Z$ production is in principle very different from
Higgs production because the $Z$ boson has a non-negligible coupling
to the light quarks. For simplicity, we will not take these couplings
into account; to avoid confusion, we refer to the $Z$ boson that
couples only with heavy quarks as $Z'$, even when we take its mass to
be equal to $91.2$~GeV as in the Standard Model.

We have calculated the 5F scheme cross sections by using a private
code~\cite{Maltoni:2005wd}, which has been cross-checked at LO and NLO
against \amc.  The 4F scheme cross section has been computed with \amc.  Our
settings are the same as in the Higgs production computation. We take
the same value $\mu$ for the factorisation and renormalisation scales.

\begin{figure}[ht]
\centering
\includegraphics[width=0.45\textwidth]{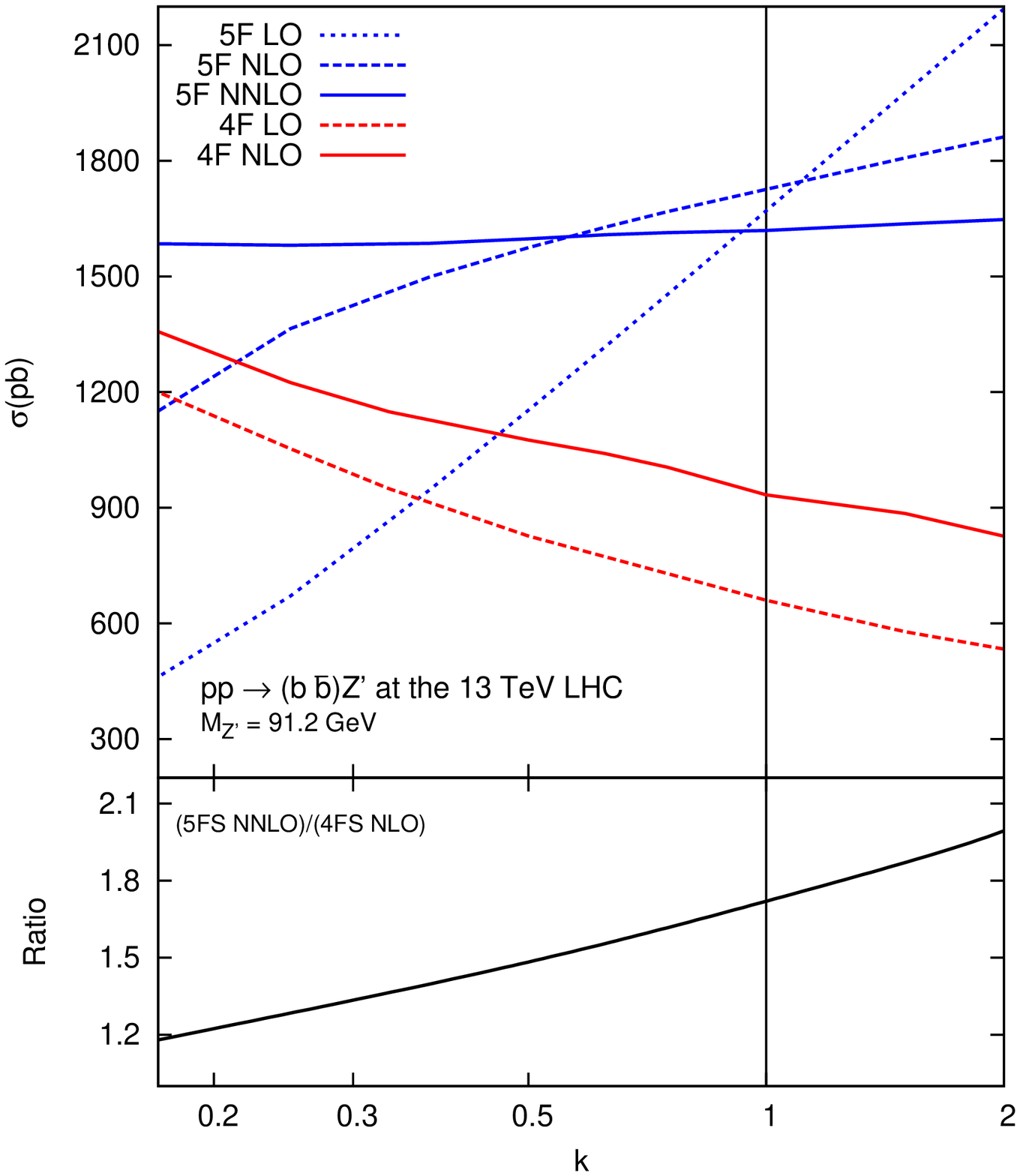}
\includegraphics[width=0.45\textwidth]{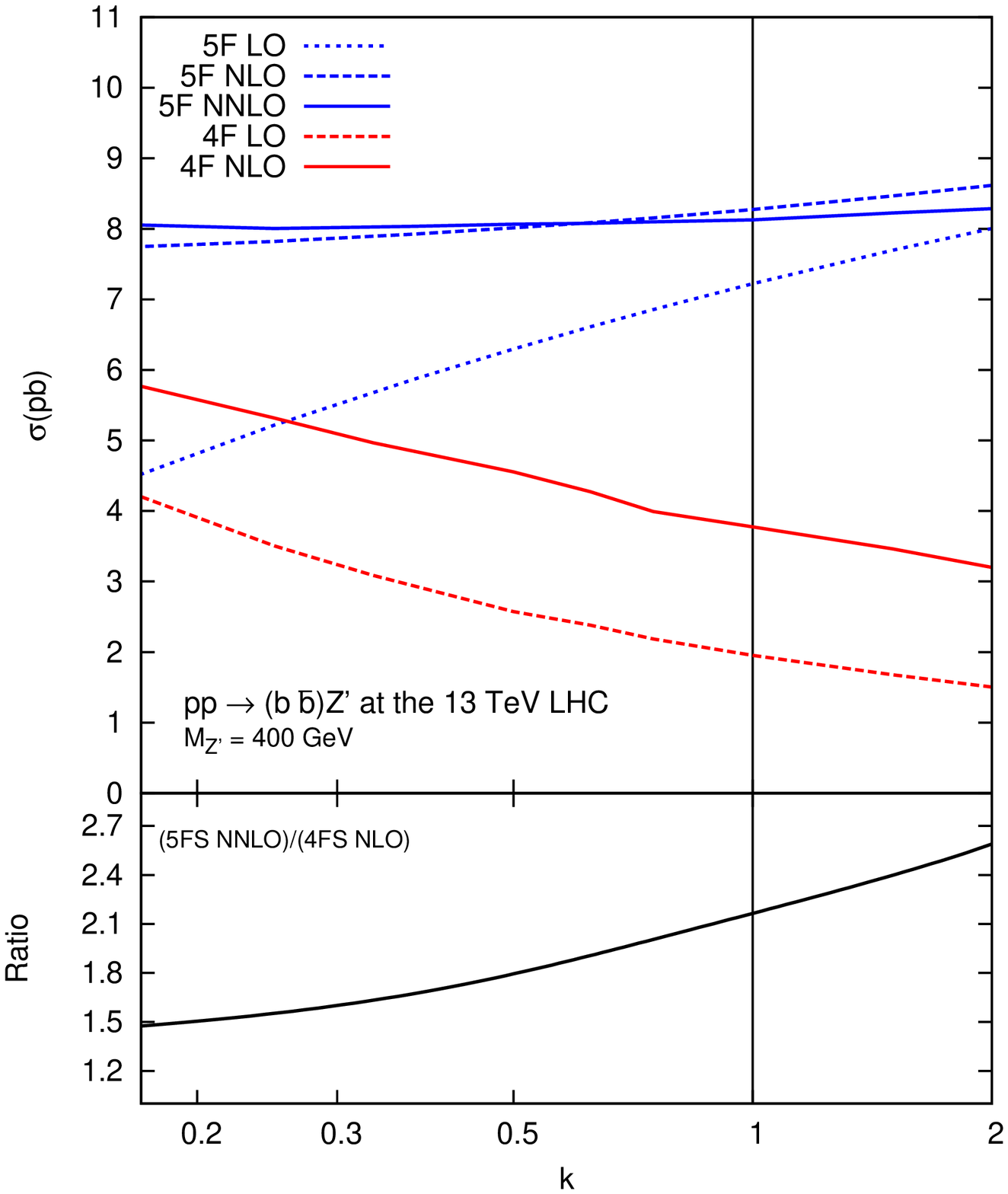}
\caption{\label{fig:bbZ} Cross sections for bottom-fusion initiated
  $Z'$ boson production in the 5F and 4F schemes for LHC 13 TeV as
  functions of $k=\mu/M_{Z'}$. $M_{Z'}=91.2$ GeV (left) and
  $M_{Z'}=400$ GeV (right).  Settings are specified in the text.
}
\end{figure}

Results are presented in Fig.~\ref{fig:bbZ} as functions of
$k=\mu/M_{Z'}$ for $M_{Z'}=91.2$ GeV and $M_{Z'}=400$ GeV
respectively.  We observe that for $\mu=M_{Z'}$ the best 5F scheme prediction
exceeds the 4F scheme prediction by almost 30\%, while their difference is
reduced at lower values of the scales.  In this respect the behaviour
of the 4F vs 5F scheme predictions reflects what we have already observed in
Fig.~\ref{fig:bbH}.  We note, however, that the scale dependence of
the 5F scheme predictions for $Zb\bar{b}$ is quite different with
respect to the $Hb\bar{b}$ when $m_H=125$ GeV.  In the case of
$Zb\bar{b}$ is quite mild already at NLO and the perturbative
expansion seems to converge more quickly for higher values of $\mu$
around $\mu=M_{Z'}$.  The behaviour of the 5F calculations for
$M_H=M_{Z'}=400$ GeV cases, on the other hand, do not show any
significant qualitative difference, apart from the fact that
$Zb\bar{b}$ results have in general a milder scale dependence.  The
different scale sensitivity (with $\mu_R=\mu_F$) of the two processes
can be traced back to the fact that while the Yukawa interaction
renormalises under QCD, the EW current (and corresponding charge) is
conserved, resulting in general in a milder scale dependence of the
$Zb\bar{b}$ predictions.

\subsection{Future Colliders}

The perspective of a proton-proton collider at a
centre--of--mass energy of 100 TeV would open up a new territory
beyond the reach of the LHC. New heavy particles associated with new
physics sector may be discovered and new interactions unveiled.  At
such large energies, essentially all SM particles can be
considered as massless, including the top quarks. We therefore
expect collinear enhancements in top-quark initiated processes.  In
Ref.~\cite{Dawson:2014pea} the question of whether the top quark should be
treated as an ordinary parton at high centre-of-mass energy, 
thereby defining a 6FNS, is scrutinised, and the impact of resumming
collinear logs of the top quark mass is assessed. This analysis is
performed in the context of charged Higgs boson production at 100 TeV.
In Ref.~\cite{Han:2014nja}, the impact of resumming initial-state
collinear logarithms in the associated heavy Higgs ($M_H>5$ TeV) and
top pair production (with un-tagged top quarks) is examined and it is
found to be very large at large Higgs masses.

\begin{figure}[ht]
\centering
\includegraphics[width=0.33\textwidth]{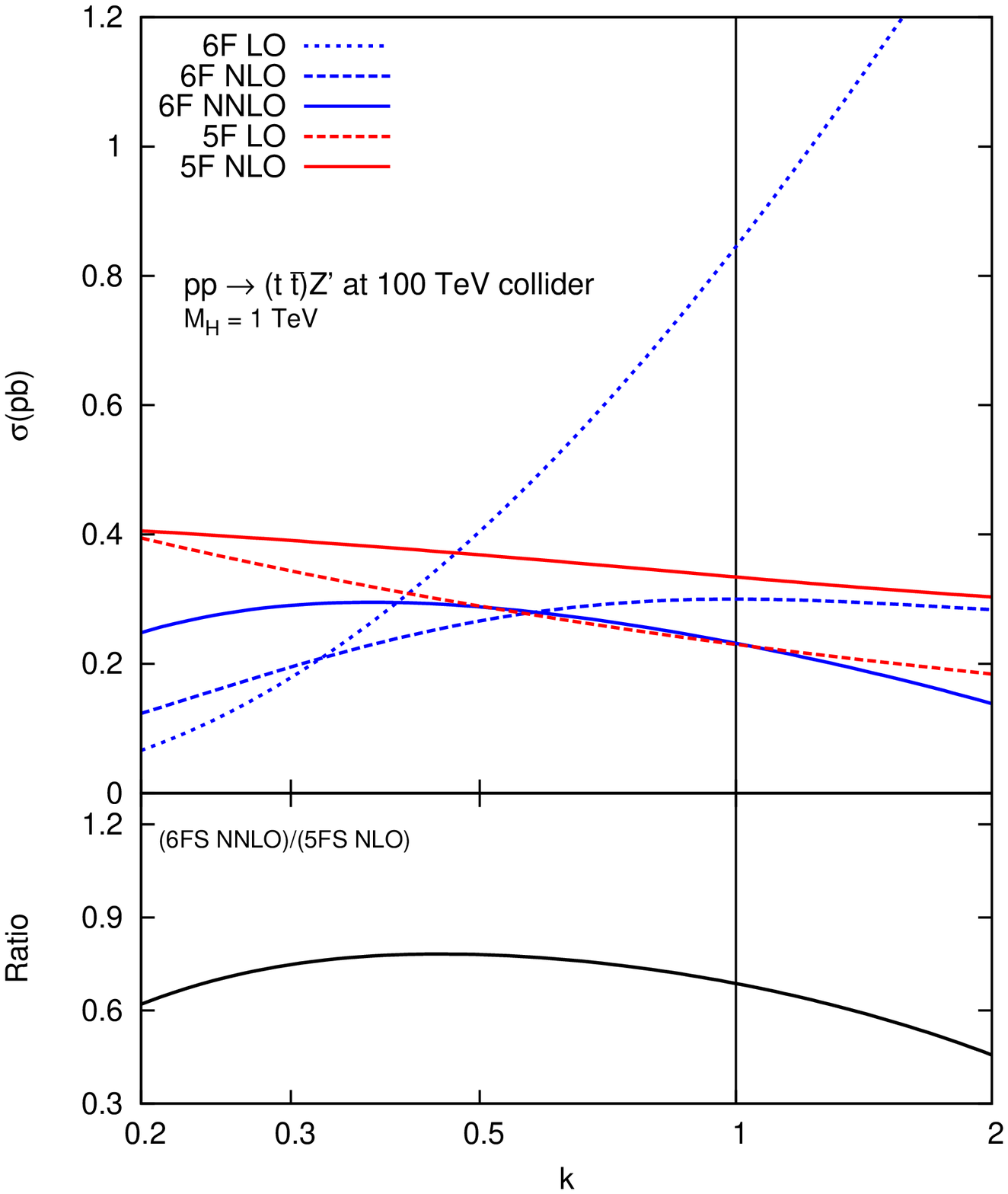}~
\includegraphics[width=0.33\textwidth]{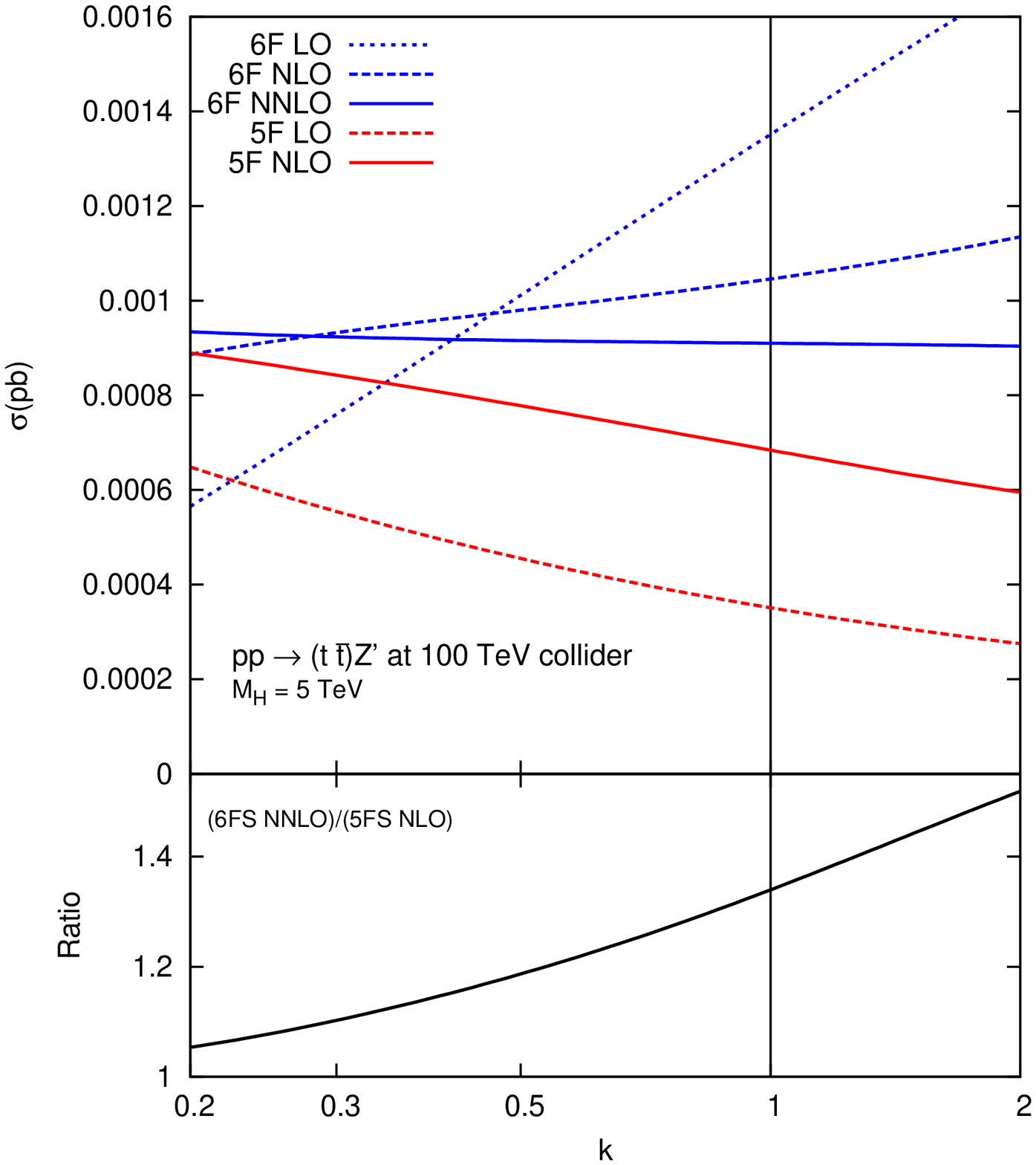}~
\includegraphics[width=0.33\textwidth]{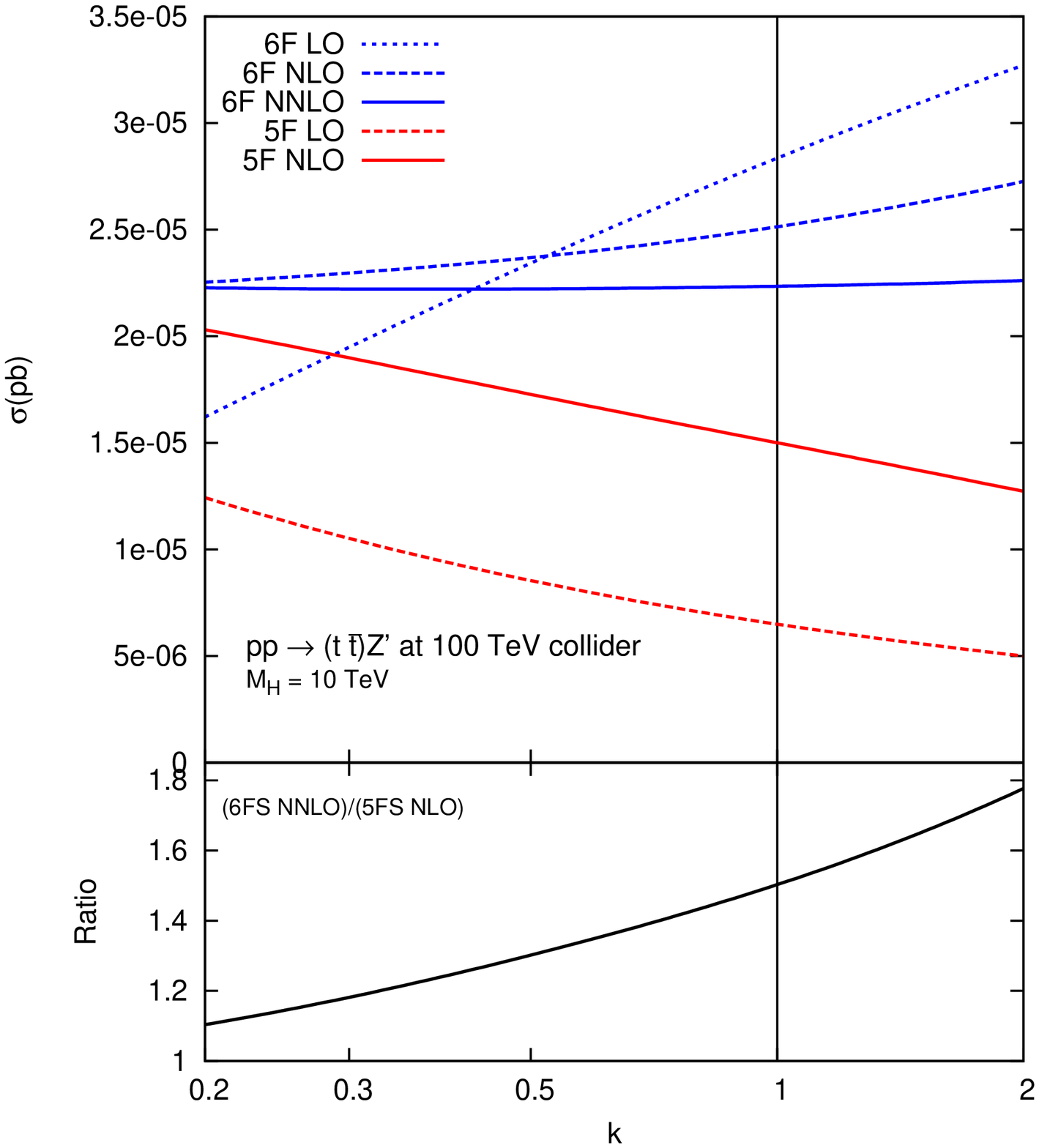}
\caption{\label{fig:ttZ} 
Cross sections for $t\bar t$ initiated $Z'$ production in the 6F and
5F schemes at a 100 TeV $pp$ collider as functions of
$k=\mu/M_{Z'}$. Top mass: $m_t= 173$ GeV.  Mass of the heavy boson:
$M_{Z'}=$ 1 TeV (left), $M_{Z'}=$ 5 TeV (centre), $M_{Z'}=$ 10 TeV
(right). The inlay below shows the ratio of the cross sections in the
6F and 5F schemes. }
\end{figure}

\begin{figure}[ht] \centering
\includegraphics[width=0.5\textwidth]{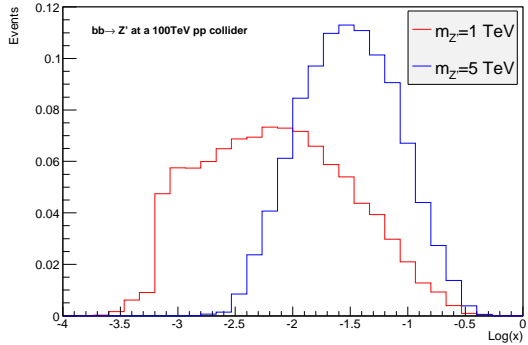} 
\caption{\label{fig:ZPdist} Normalised distribution of momentum
  fraction $x$ carried by the $t\bar t$ initiated $Z'$ production in
  the 6F scheme distributions at LO in a 5F scheme for $M_{Z'}=1$ TeV
  and $M_{Z'} = 5$ TeV at a 100 TeV collider.  Events were generated
  at values of the scales $\mu_R=\mu_F=H_T/4$.  Input PDF: NNPDF30 LO
  $n_f=5$ ($\alpha_s(M_Z)=0.130$).}
\end{figure}

In Fig.~\ref{fig:ttZ} the total cross sections for the production of a
$Z'$ boson of mass $M_{Z'}=$ 1 TeV (left), $M_{Z'}=$ 5 TeV (centre), 
$M_{Z'}=$ 10 TeV (right) are plotted in the 5F and
6F schemes as a function of the renormalisation and
factorisation scales, which are identified and varied between $0.2
M_{Z'}$ and $2M_{Z'}$.  Results are obtained by using \amc\ for the 5F scheme 
and a private code for the 6F scheme. Results in the 6F scheme have been cross-
checked up to NLO against \amc. We have set $m_t^{\rm pole}=172.5$ GeV and
turned off the coupling of the $Z'$ heavy boson to all lighter
quarks.
Firstly, we observe that the $M_{Z'}=$ 1 TeV case is quite different from
the $M_{Z'}=$ 5 TeV and $M_{Z'}=$ 10 TeV, which in turn display a very
similar pattern to the $b$ initiated processes with similar
$m_Q/M_{Z'}$ and $ M_{Z'}/\sqrt{s}$ ratios. The behaviour of the
leading-order cross section in the 6F scheme for $M_{Z'}=$ 1 TeV is
mitigated at higher masses and at higher orders (NLO).  At NNLO the
6F-scheme cross section displays a similar scale dependence as the NLO
cross section in the 5F scheme with a residual difference of about
40\% between the two best predictions in the two schemes. To further
investigate these differences, in Fig.~\ref{fig:ZPdist} we plot the
distribution of the fraction of momentum carried by the top quarks for
$M_{Z'}=$ 1 TeV and $M_{Z'}=$ 5 TeV in the 6F schemes.
As expected, compared to heavier masses, the production of a $M_{Z'}=$ 1
TeV happens mostly at threshold and it is dominated by smaller values
of Bjorken $x$.  The ratio $M_{Z'}/m_t \simeq 6 $ is not very large to
start with (for comparison $M_{Z}/m_b \simeq 20 $) and initial-state
quark collinear configurations are not dominant. We conclude that in the
$M_{Z'}=$ 1 TeV case the differences between the two schemes are to be
associated to the absence of power-like mass terms in the 6F
calculation.

\section{Conclusions}
\label{sec:conclusions}

In this work we have considered the use of four- and five-flavour
schemes in precision physics at the LHC and in the context of
$b$-initiated Higgs and $Z$ production. We have extended previous work
done for processes involving a single $b$ quark in the initial state
to cases in which two are present. We have followed a
``deconstructing'' methodology where the impacts of the various
sources of differences between the schemes have been evaluated one by
one.

Firstly, we have obtained the form of the collinear logarithms in the
four-flavour scheme by performing the explicit computation of the $2\to
3$ body scattering process and studying the collinear limit using as
natural variables the $t$-channel invariants. We have then compared
the resulting expression with the corresponding cross section in the
5-flavor scheme as calculated by only keeping the explicit log in the
$b$-quark PDF, i.e. without resummation.  This has allowed us to
assess the analytic form and therefore the size of the collinear
logarithms and to propose a simple procedure to identify the relevant
scales in the processes where the results in the two schemes should be
evaluated and compared.  In so doing we have considered cases where
power-like effects in the mass of the heavy quarks were assumed (and
then checked a posteriori by comparing to the full result)
unimportant. Secondly, we have explicitly estimated the effects of the
resummation by studying fully evolved $b$ PDF with truncated
expansions at finite order.

We have then applied our general approach to the case of Higgs and $Z$
boson production in association with $b$ quarks at the LHC and to
heavy $Z'$ production in association with top quarks at a future 100
TeV collider.  We have found that the resummation increases the cross
section in most cases by about 20\% (sometimes reaching 30\%) at the
LHC and in general leads to a better precision.  On the other hand,
the 4F scheme predictions (5F scheme in the case of associated top-quark production)
at NLO also display a consistent perturbative behaviour when evaluated
at suitable scales. They should therefore should be used when the
heavy-quark mass effects are not negligible and to predict
distributions involving the heavy quarks in the final state.

\section*{Acknowledgements}
We would like to thank Stefano Forte, Paolo Nason, Alex Mitov and
Davide Napoletano for many useful discussions on this topic and for
comments on this work. In particular we thank Davide Napoletano for
providing the code that we used to check the effects of the inclusion
of higher order logs in the NLO five-flavor scheme cross sections.  
We thank the
Kavli Institute for Theoretical Physics in Santa Barbara for hosting
the authors during the completion of this manuscript.  This research
was supported in part by the National Science Foundation under Grant
No. NSF PHY11-25915.  The work of G.R. is supported in part by an
Italian PRIN2010 grant.  

\appendix
\section{Cross section in the collinear limit}
\label{appxs}

\begin{figure}[htb]
\includegraphics[width=0.22\textwidth]{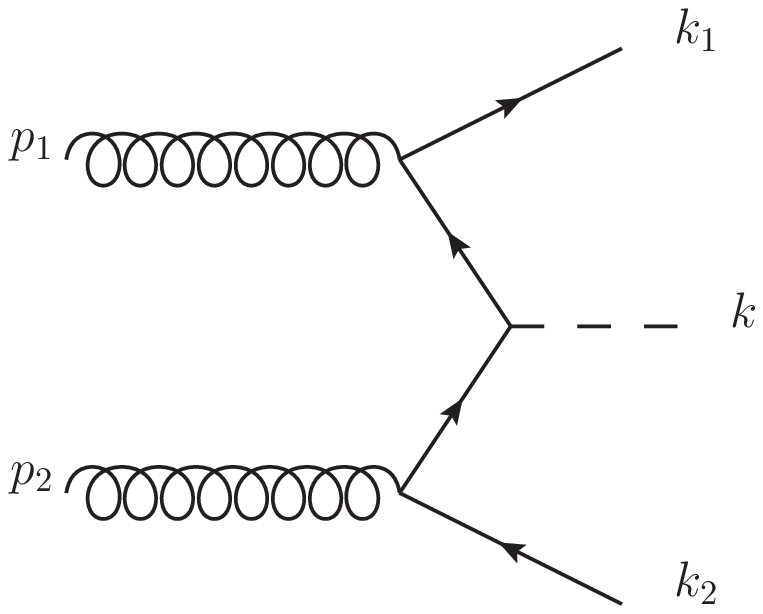}
\includegraphics[width=0.22\textwidth]{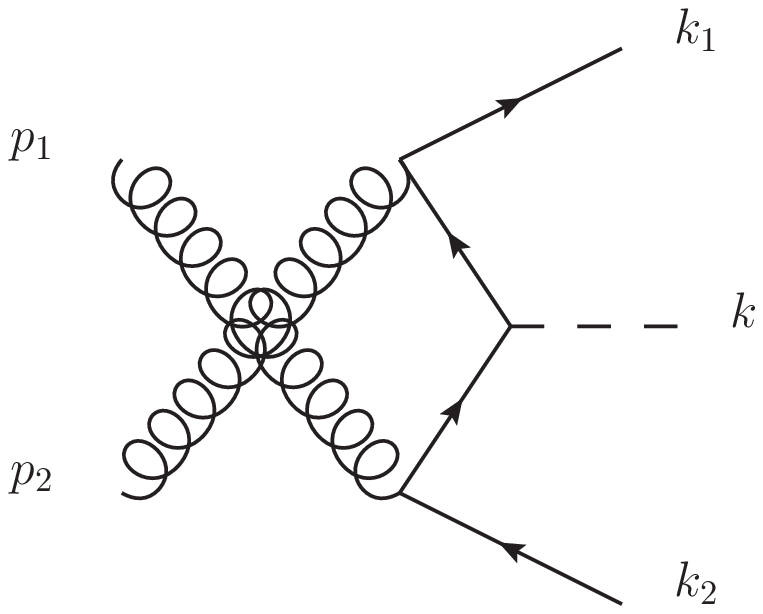}
\includegraphics[width=0.22\textwidth]{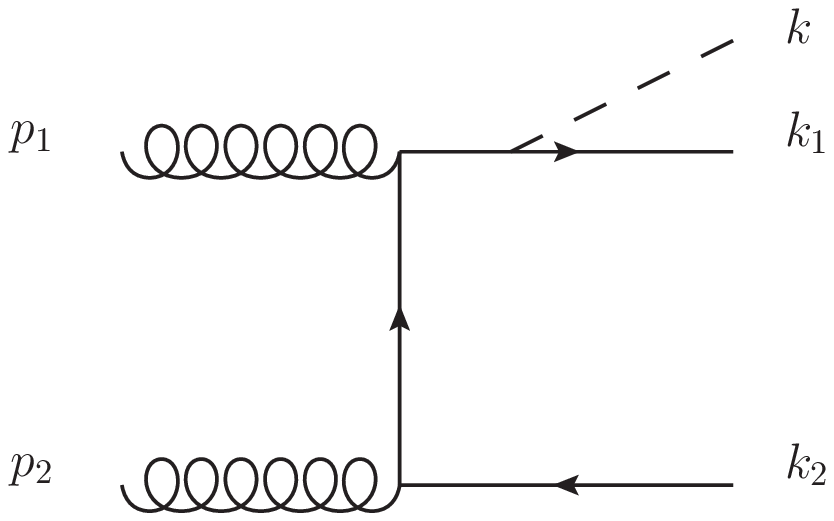}
\includegraphics[width=0.22\textwidth]{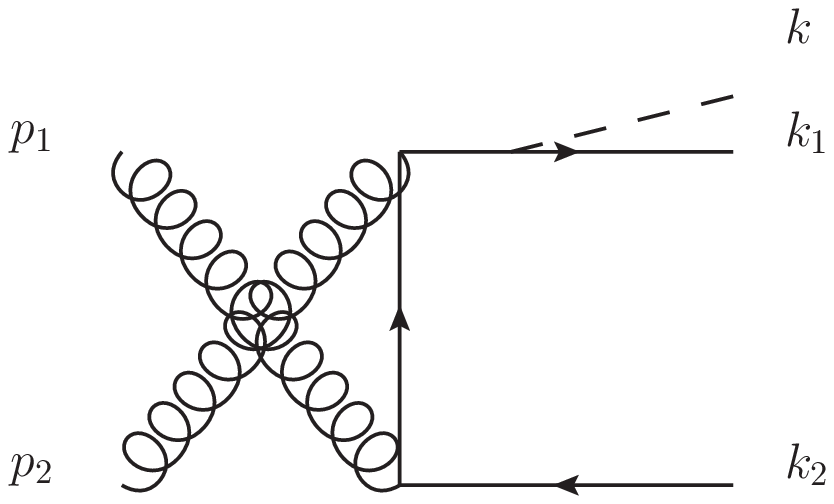}\\
\includegraphics[width=0.22\textwidth]{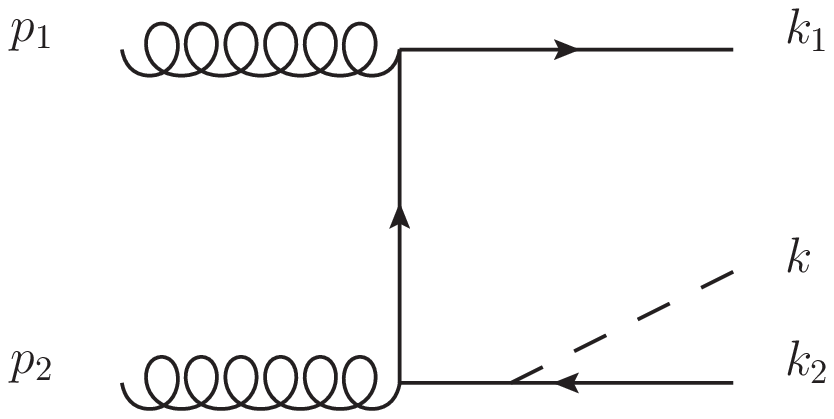}
\includegraphics[width=0.22\textwidth]{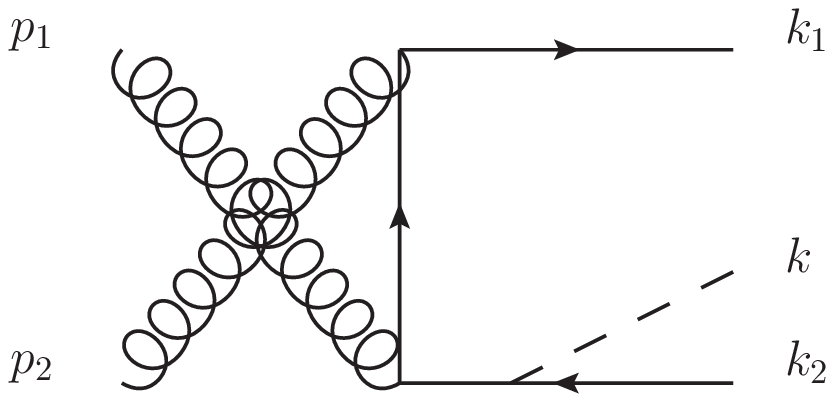}
\includegraphics[width=0.22\textwidth]{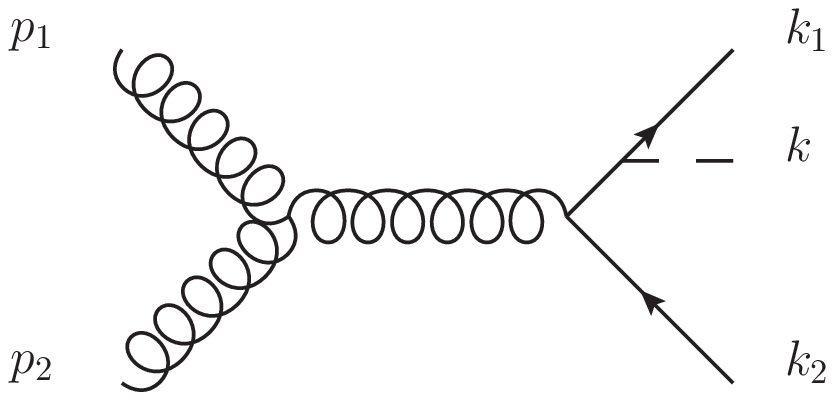}
\includegraphics[width=0.22\textwidth]{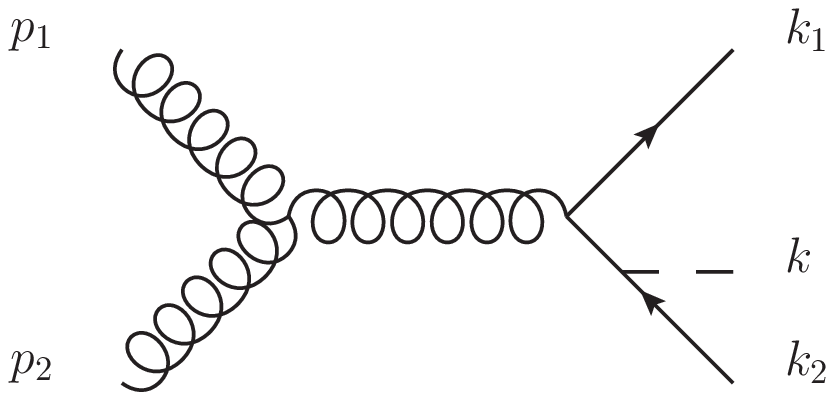}
\caption{\label{fig:feynLO}Leading order diagrams for $gg\to b\bar b H$.}
\end{figure}

In this Appendix we illustrate in some detail the calculation
of the cross section for the partonic process
\begin{equation}
g(p_1)+g(p_2)\to b(k_1)+\bar b(k_2)+H(k)
\end{equation}
in the limit of collinear emission of $b$ quarks.
We choose, as independent kinematic invariants,
\begin{eqnarray}
\hat s &=& (p_1+p_2)^2=2p_1p_2\\
t_1 &=& (p_1-k_1)^2=-2p_1k_1+m_b^2\\
t_2 &=& (p_2-k_2)^2=-2p_2k_2+m_b^2\\
s_1 &=& (k_1+k)^2   = 2k_1k + m_b^2 + M_H^2\\
s_2 &=& (k_2+k)^2  = 2k_2k + m_b^2 + M_H^2.
\end{eqnarray} 
The remaining invariants
\begin{align}
u_1&=(p_1-k_2)^2=-2p_1k_2+m_b^2\\
u_2&=(p_2-k_1)^2=-2p_2k_1+m_b^2\\
s_{12}&=(k_1+k_2)^2=2k_1k_2+2m_b^2\\
t&=(p_1-k)^2-M_H^2=-2kp_1\\
u&=(p_2-k)^2-M_H^2=-2kp_2
\end{align}
are related to the independent invariants by
\begin{align}
&u_1=s_1-\hat s-t_2+m_b^2\label{u1}\\
&u_2=s_2-\hat s-t_1+m_b^2\label{u2}\\
&t=-s_1+t_2-t_1+m_b^2\\
&u=-s_2+t_1-t_2+m_b^2\\
&s_{12}=\hat s-s_1-s_2+M_H^2+2m_b^2.
\end{align}
The leading-order Feynman diagrams are shown in 
Fig.~\ref{fig:feynLO}. 
The squared invariant amplitude (averaged over initial state summed over
final state spin and color variables) has the general structure
\begin{equation}
\left|\mathcal{M}\right|^2
=\frac{G(s,s_1,s_2,t_1,t_2)}
{(t_1-m_b^2)^2(t_2-m_b^2)^2(u_1-m_b^2)^2(u_2-m_b^2)^2}.
\label{squaredM}
\end{equation}
The function $G(s,s_1,s_2,t_1,t_2)$ is a polynomial in $t_1,t_2$.
It can be shown on general grounds~\cite{Keller:1998tf,Catani:2000ef}
that each double pole is suppressed by a factor of $m_b^2$.
Furthermore, it is well known that
collinear singularities do not arise in interference terms
among different amplitudes. Thus, 
\begin{equation}
\left|\mathcal{M}\right|^2
=\frac{G_t}{(t_1-m_b^2)(t_2-m_b^2)}
+\frac{G_u}{(u_1-m_b^2)(u_2-m_b^2)}+\left|\mathcal{M}\right|^2_{\rm reg}
\label{squaredM0}
\end{equation}
where the term $\left|\mathcal{M}\right|^2_{\rm reg}$
does not give rise to collinear singularities in the limit $m_b=0$.
An explicit calculation gives
\begin{equation}
G_t=G_u=\frac{32\alpha_s^2\pi^2 m_b^2 G_F M_H^2\sqrt{2}}{3}
\frac{P_{qg}(z_1)}{z_1}\frac{P_{qg}(z_2)}{z_2},
\end{equation}
where
\begin{equation}
z_1=\frac{M_H^2}{s_1};\qquad
z_2=\frac{M_H^2}{s_2}
\end{equation}
and $P_{qg}(z)$ is defined in Eq.~(\ref{Pqg}).

The 3-body phase-space invariant measure
\begin{align}
&d\phi_3(p_1,p_2;k_1,k_2,k)
\nonumber\\
&=\frac{d^3k_1}{(2\pi)^32k_1^0}\frac{d^3k_2}{(2\pi)^32k_2^0}
\frac{d^3k}{(2\pi)^32k^0}(2\pi)^4\delta(p_1+p_2-k_1-k_2-k)
\end{align}
can be factorised as
\begin{equation}
d\phi_3(p_1,p_2;k_1,k_2,k)=\frac{dt_1}{2\pi}\frac{dt_2}{2\pi}
d\phi_2(p_1;k_1,q_1)d\phi_2(p_2;k_2,q_2)d\phi_1(q_1,q_2;k),
\end{equation}
where
\begin{equation}
q_1^2=t_1;\qquad q_2^2=t_2.
\end{equation}
We now compute each factor explicitly. We have
\begin{align}
d\phi_2(p_1;k_1,q_1)
&=\frac{d^3k_1}{(2\pi)^32k_1^0}\frac{d^3q_1}{(2\pi)^32q_1^0}
(2\pi)^4\delta(p_1-k_1-q_1)
\nonumber\\
&=\frac{1}{16\pi^2}
\frac{|\vec k_1|^2d|\vec k_1|d\cos\theta_1d\phi_1}{k_1^0q_1^0}
\delta(p_1^0-k_1^0-q_1^0)
\end{align}
where
\begin{align}
&k_1^0=\sqrt{|\vec k_1|^2+m_b^2}
\\
&q_1^0=\sqrt{{|\vec p_1|}^2+|\vec k_1|^2-2|\vec p_1||\vec k_1|\cos\theta_1+t_1}
\end{align}
We may now integrate over $\cos\theta_1$
using the delta function
\begin{equation}
\delta(p_1^0-k_1^0-q_1^0)=\frac{q_1^0}{|\vec p_1||\vec k_1|}
\delta(\cos\theta_1-\cos\bar\theta_1)
\end{equation}
with $\bar\theta_1$ a solution of
\begin{equation}
p_1^0-\sqrt{|\vec k_1|^2+m_b^2}
-\sqrt{|\vec p_1|^2+|\vec k_1|^2-2|\vec p_1||\vec k_1|\cos\bar\theta_1+t_1}=0.
\label{t1gen}
\end{equation}
This gives
\begin{equation}
d\phi_2(p_1;k_1,q_1)
=\frac{1}{16\pi^2}\frac{|\vec k_1|d|\vec k_1|d\varphi_1}{k_1^0|\vec p_1|};
\qquad
d\phi_2(p_2;k_2,q_2)
=\frac{1}{16\pi^2}\frac{|\vec k_2|d|\vec k_2|d\varphi_2}{k_2^0|\vec p_2|}
\end{equation}
and therefore
\begin{equation}
d\phi_3(p_1,p_2;k_1,k_2,k)=\frac{1}{1024\pi^6}dt_1 dt_2
\frac{|\vec k_1|d|\vec k_1|d\varphi_1}{k_1^0|\vec p_1|}
\frac{|\vec k_2|d|\vec k_2|d\varphi_2}{k_2^0|\vec p_2|}
d\phi_1(q_1,q_2;k).
\end{equation}
It will be convenient to adopt the centre-of-mass frame,
where
\begin{equation}
p_1=\frac{\sqrt{\hat s}}{2}(1,0,0,1),\qquad
p_2=\frac{\sqrt{\hat s}}{2}(1,0,0,-1)
\label{p1p2}
\end{equation}
In this frame
\begin{align}
&s_1=(k+k_1)^2=(p_1+p_2-k_2)^2=\hat s+m_b^2
-2\sqrt{\hat s}\sqrt{|\vec k_2|^2+m_b^2}
\label{s1}
\\
&s_2=(k+k_2)^2=(p_1+p_2-k_1)^2=\hat s+m_b^2
-2\sqrt{\hat s}\sqrt{|\vec k_1|^2+m_b^2}
\label{s2}
\end{align}
and therefore
\begin{equation}
\frac{|\vec k_1|d|\vec k_1|}{k_1^0|\vec p_1|}
\frac{|\vec k_2|d|\vec k_2|}{k_2^0|\vec p_2|}
=\frac{ds_1}{\hat s}\frac{ds_2}{\hat s}.
\end{equation}
Furthermore, we may use the invariance of the cross
section upon rotations about the $z$ axis to replace
\begin{equation}
d\varphi_1d\varphi_2\to 2\pi d\varphi;\qquad \varphi=\varphi_1-\varphi_2.
\end{equation}
Finally,
\begin{equation}
d\phi_1(q_1,q_2;k)=2\pi \delta\left((q_1+q_2)^2-M_H^2\right),
\end{equation}
and therefore
\begin{equation}
d\phi_3(p_1,p_2;k_1,k_2,k)
=\frac{1}{256\pi^4\hat s^2}ds_1 ds_2 dt_1 dt_2\,
d\varphi\delta\left((q_1+q_2)^2-M_H^2\right).
\end{equation}
It is a tedious, but straightforward, task
to show that, upon integration over the azimuth $\varphi$ using the delta
function, this expression is the same as the one given 
in~\cite{Byckling:1971vca}
for the three-body phase-space measure in terms of four invariants.

The two invariants $u_1,u_2$ are related to independent
invariants through Eqs.~(\ref{u1},\ref{u2}), which can be written
\begin{align}
&u_1-m_b^2=-(t_2-a_2)
\\
&u_2-m_b^2=-(t_1-a_1)
\end{align}
where we have defined
\begin{equation}
a_1=s_2-\hat s;\qquad a_2=s_1-\hat s.
\end{equation}
The bounds for $t_1$ are easily obtained. In the
centre-of-mass frame we have
\begin{align}
t_1=&\frac{1}{2}\left[a_1+m_b^2-\cos\bar\theta_1
\sqrt{(a_1+m_b^2)^2-4m_b^2(a_1+\hat s)}
\right]\\
t_2=&\frac{1}{2}\left[a_2+m_b^2+\cos\bar\theta_2
\sqrt{(a_2+m_b^2)^2-4m_b^2(a_2+\hat s)}
\right].
\end{align}
The upper and lower bound are obtained for
$\cos\bar\theta_1=\pm 1$, $\cos\bar\theta_2=\pm 1$.
We get
\begin{equation}
t_1^-\leq t_1\leq t_1^+;\qquad
t_2^-\leq t_2\leq t_2^+,
\label{t12bounds}
\end{equation}
where
\begin{align}
&t_1^\pm=\frac{1}{2}\left[a_1+m_b^2\pm\sqrt{(a_1+m_b^2)^2
-4m_b^2(a_1+\hat s)}\right]
\label{t1bounds}
\\
&t_2^\pm=\frac{1}{2}\left[a_2+m_b^2\pm\sqrt{(a_2+m_b^2)^2
-4m_b^2(a_2+\hat s)}\right].
\label{t2bounds}
\end{align}
For small $m_b^2$,
\begin{equation}
t_i^+=m_b^2+\frac{m_b^2\hat s}{a_i}+O(m^4);\qquad 
t_i^-=a_i-\frac{m_b^2\hat s}{a_i}+O(m^4);\qquad i=1,2.
\end{equation}

All the ingredients to compute the total partonic cross section in the
collinear limit are now available. In this limit, the relative azimuth
$\phi$ between $b$ and $\bar b$ is irrelevant, and simply provides a factor
of $2\pi$. Furthermore
\begin{equation}
\hat s=\frac{M_H^2}{z_1z_2};\qquad s_1=\hat s z_2;\qquad s_2=\hat s z_1
\end{equation}
and therefore
\begin{equation}
\frac{ds_1\,ds_2}{\hat s^2}=dz_1\,dz_2.
\end{equation}
The integrals over $t_1,t_2$ are easily computed:
\begin{align}
&\int_{t_i^-}^{t_i^+}dt_i\,\frac{1}{t_i-m_b^2}
=\log\frac{a_1^2}{m_b^2\hat s}+O(1)
=\log\frac{M_H^2}{m_b^2}\frac{(1-z_i)^2}{z_1z_2}
\\
&\int_{t_i^-}^{t_i^+}dt_i\,\frac{1}{t_i-a_i}
=-\log\frac{a_i^2}{m_b^2\hat s}+O(1)
=-\log\frac{M_H^2}{m_b^2}\frac{(1-z_i)^2}{z_1z_2}+O(1).
\end{align}
Finally,
\begin{equation}
\delta\left((q_1+q_2)^2-M_H^2\right)=\delta(z_1z_2\hat s-M_H^2).
\end{equation}
We find
\begin{align}
\hat\sigma^{\rm 4F,coll}(\hat \tau)
&=\frac{1}{2\hat s}\int d\phi_3(p_1,p_2;k_1,k_2,k)\,G_u
\left[\frac{1}{(t_1-m_b^2)(t_2-m_b^2)}+\frac{1}{(t_1-a_1)(t_2-a_2)}\right]
\nonumber\\
&=\hat\tau\frac{\alpha_s^2}{4\pi^2}
\frac{m_b^2}{M_H^2} \frac{G_F\pi}{3\sqrt{2}}
2\int_0^1dz_1\int_0^1dz_2\,\delta(z_1z_2-\hat\tau)
\nonumber\\
&
\times P_{qg}(z_1)\log\left[\frac{M_H^2}{m_b^2}\frac{(1-z_1)^2}{\hat\tau}\right]
P_{qg}(z_2)\log\left[\frac{M_H^2}{m_b^2}\frac{(1-z_2)^2}{\hat\tau}\right].
\end{align}

\renewcommand{\em}{}
\bibliographystyle{UTPstyle}
\bibliography{bbH-JHEP}

\end{document}